\documentclass{article}
\usepackage{arxiv}
\usepackage{graphicx}
\usepackage{subcaption}
\usepackage{bm}
\usepackage{amsmath}
\usepackage{amsfonts}
\usepackage{amssymb}
\usepackage{amsthm}
\usepackage[ruled,vlined]{algorithm2e}
\usepackage[noend]{algpseudocode}

\usepackage{appendix}
\usepackage{color}

\newcommand{\Aref}[1]{Alg.\,\ref{#1}}

\newcommand{\aref}[1]{App.\,\ref{#1}}
\newcommand{\fref}[1]{Fig.\,\ref{#1}}
\newcommand{\tref}[1]{Table\,\ref{#1}}
\newcommand{\eref}[1]{Eq.\,(\ref{#1})}

\newcommand{\sref}[1]{Sec.\!~\ref{#1}}

\newcommand{\cref}[1]{Ref.\,\cite{#1}}
\newcommand{\crefs}[1]{Refs.\,\cite{#1}}

\newcommand{\etal}{{\it et al.} }

\newcommand{\rotation}{\mathbf{Q}}
\newcommand{\reflection}{\mathbf{R}}

\newcommand{\tighteq}{\mkern1.5mu{=}\mkern1.5mu}

\newcommand{\Abb}{\mathbb{A}}
\newcommand{\Ibb}{\mathbb{I}}
\newcommand{\Jbb}{\mathbb{J}}
\newcommand{\Lbb}{\mathbb{L}}
\newcommand{\Kbb}{\mathbb{K}}
\newcommand{\Cbb}{\mathbb{C}}
\newcommand{\Dbb}{\mathbb{D}}

\newcommand{\Obb}{\mathbb{O}}

\newcommand{\ab}{\mathbf{a}}

\newcommand{\cb}{\mathbf{c}}

\newcommand{\eb}{\mathbf{e}}

\newcommand{\pb}{\mathbf{p}}

\newcommand{\nb}{\mathbf{n}}

\newcommand{\Ab}{\mathbf{A}}
\newcommand{\Bb}{\mathbf{B}}
\newcommand{\Cb}{\mathbf{C}}

\newcommand{\Eb}{\mathbf{E}}

\newcommand{\Gb}{\mathbf{G}}
\newcommand{\Qb}{\mathbf{Q}}

\newcommand{\Nb}{\mathbf{N}}

\newcommand{\Ib}{\mathbf{I}}

\newcommand{\Sb}{\mathbf{S}}

\newcommand{\Ac}{\mathcal{A}}

\newcommand{\Ic}{\mathcal{I}}
\newcommand{\Lc}{\mathcal{L}}

\newcommand{\bs}{\mathsf{b}}
\newcommand{\cs}{\mathsf{c}}

\newcommand{\vs}{\mathsf{v}}

\newcommand{\As}{\mathsf{A}}

\newcommand{\Ds}{\mathsf{D}}

\newcommand{\Gs}{\mathsf{G}}

\newcommand{\Ms}{\mathsf{M}}

\newcommand{\Us}{\mathsf{U}}
\newcommand{\Vs}{\mathsf{V}}

\newcommand{\epsilonb}{{\boldsymbol{\epsilonb}}}
\newcommand{\alphab}{{\boldsymbol{\alpha}}}

\newcommand{\varepsilonb}{{\boldsymbol{\varepsilon}}}
\newcommand{\tr}{{\operatorname{tr}}}
\newcommand{\perm}{{\operatorname{perm}}}

\newcommand{\partialb}{{{\boldsymbol{\partial}}}}
\newcommand{\NN}{{\mathsf{N}\!\mathsf{N}}}

\newcommand{\basis}{{\mathcal{B}}}
\newcommand{\invariants}{{\mathcal{I}}}
\newcommand{\RE}{{R}}

\usepackage{todonotes}

\usepackage{afterpage}
\usepackage{longtable}

\usepackage{bbm}

\newcommand{\structuraltensor}{\mathbb{A}}
\newcommand{\strain}{\mathbf{E}}

\newcommand{\stress}{\mathbf{S}}
\newcommand{\grp}{\mathcal{G}}
\newcommand{\generator}{\mathbf{G}}
\renewcommand{\RE}{\mathbbmss{R}}
\renewcommand{\AA}{\mathbbmss{A}}
\newcommand{\VV}{\mathbbmss{V}}
\newcommand{\WV}{\mathbbmss{W}}
\renewcommand{\perm}{\boldsymbol{\varepsilon}}
\renewcommand{\perm}{\boldsymbol{\varepsilon}}
\newcommand{\SOthree}{SO(3)}
\newcommand{\Othree}{O(3)}

\title{{\bf A General, Automated Method for Building Structural Tensors of Arbitrary Order for Anisotropic Function Representations}}

\renewcommand{\shorttitle}{An Automated Method for Building Structural Tensors of Arbitrary Order}

\author{
Ravi G. Patel,\thanks{equal contribution} \\
{\it Sandia National Laboratories},\\
Albuquerque, NM 87185
\And
Reese E. Jones$^*$\thanks{{correspondence: \tt rjones@sandia.gov}}, \\
{\it Sandia National Laboratories},\\
Livermore, CA 94551
\And
Brian N. Granzow, \\
{\it Sandia National Laboratories},\\
Albuquerque, NM 87185
\And
D. Thomas Seidl, \\
{\it Sandia National Laboratories},\\
Albuquerque, NM 87185
\And
Jan N. Fuhg, \\
{\it
Department of Aerospace Engineering \& Engineering Mechanics}, \\
{\it
\& The Oden Institute of Computational Science and Engineering,} \\
{\it
The University of Texas at Austin}, \\
Austin, TX 78712
}
\date{}

\begin{document}

\maketitle

\begin{abstract}
We present a general, constructive procedure to find the basis for tensors of arbitrary order subject to linear constraints by transforming the problem to that of finding the nullspace of a linear operator.
The proposed method utilizes standard numerical linear algebra techniques that are highly optimized and well-behaved.
Our primary applications are in mechanics where modulus tensors and so-called \emph{structure tensors} can be used to characterize anisotropy of functional dependencies on other inputs such as strain.
Like modulus tensors, structure tensors are defined by their invariance to transformations by symmetry group generators but have more general applicability.
The fully automated method is an alternative to classical, more intuition-reliant methods such as the Pipkin-Rivlin polynomial integrity basis construction.
We demonstrate the utility of the procedure by: (a) enumerating elastic modulus tensors for common symmetries, and (b) finding the lowest-order structure tensors that can represent all common point groups/crystal classes.
Furthermore, we employ these results in two calibration problems using neural network models following classical function representation theory: (a) learning the symmetry class and orientation of a hyperelastic material given stress-strain data, and (b) representing strain-dependent anisotropy of the stress response of a soft matrix-stiff fiber composite in a sequence of uniaxial loadings.
These two examples demonstrate the utility of the method in model selection and calibration by: (a) determining structural tensors of a selected order across multiple symmetry groups, and (b) determining a basis for a given group that allows the characterization of all subgroups.
Using a common order in both cases allows sparse regression to operate on a common function representation to select the best-fit symmetry group for the data.
\end{abstract}

\section{Introduction}

The task of finding tensors that are characteristic of a variety of physical properties is a common problem spanning quantum mechanics \cite{biedenharn1984angular}, fluid mechanics \cite{pope1975more}, solid mechanics \cite{itskov2000theory,itin2013constitutive},
material physics \cite{wallace1972thermodynamics,weiner2012statistical}, electromagnetics \cite{barrett2022topological}, and coupled physics such as piezoelectricity and magnetostriction \cite{tarn2013hamiltonian,mason1951phenomenological}.
These tensors are used to represent material properties directly and are also embedded in response functions as inputs.
Some well-known tensors of this type are: the fourth and higher order elastic moduli \cite{wallace1970thermoelastic,hiki1981higher,weiner2012statistical, d2024representation,clayton2025symmetries}, and the high order electromagnetic-deformation coupling tensors  \cite{mason1951phenomenological}.
Likewise, anisotropic response functions are well-known in elasticity and plasticity \cite{ogden2003nonlinear,cazacu2018new}.
The current procedure for function construction in mechanics, for instance, is:
(a) presume input arguments and symmetry group,
(b) find a suitable tensor that is characteristic of the required symmetries,
(c) form joint invariants (and basis) from the inputs and characteristic tensor,
(d) combine these in an isotropic function representation.
Steps (b-d) are generally a matter of searching the literature for appropriate forms that can be adapted to the particular application.
The tensors that are characteristic of symmetries due to the action of orthogonal transformations (\emph{generators}) are typically called anisotropy tensors, structure tensors, or \emph{structural} tensors.
Although the results of decades of work have been tabulated, as in Zheng's monograph \cite{zheng1993representations},
the literature is by no means complete in terms of tensors of specific order and symmetries.
The methods used to arrive at these results for the most part require significant intuition.
The method we propose just requires: (a) the generators defining the group, (b) the target order of the tensor, and (c) basic linear algebra.

In mechanics and other fields, many physical restrictions, including symmetries, are framed in terms of linear constraints on a tensor of a given order.
Unlike previous approaches for finding tensors with specialized properties, our method is efficient, automated, and uses common numerical linear algebra algorithms.
The method has a multitude of applications.
In this work, we explore: (a) finding the components of arbitrary order moduli-like tensors subject to rotational and transpositional symmetries and (b) extracting structural tensors of all the common symmetry groups for use in general, complete response function representations.
The method naturally results in a \emph{basis} for all tensors invariant to group action; hence, any/all structural tensors can be formed from this basis.
Each of these tensors is sufficient to characterize the symmetry group, unlike some representations where multiple tensors are needed to fully characterize the group \cite{zheng1993representations}.
Furthermore, we determine what tensor order is sufficient to represent all crystal symmetry groups.

In the next section, \sref{sec:related}, we put the proposed method in the context of the long history of previous work.
Then, in \sref{sec:method}, we describe the proposed method in rigorous mathematical detail.
This section also outlines the simple algorithms that enable the construction of characteristic tensors of a selected order that satisfy symmetry and other constraints.
The construction algorithms, \Aref{alg:alg} and \Aref{alg:alg2}, as well as the simplification algorithm, \Aref{alg:simplify}, lead directly to implementation in any language with access to basic numerical linear algebra.
Also \sref{sec:method} provides concise, practical applications of the algorithms.
In \sref{sec:mechanics} we apply the method to aspects of material symmetry in solid mechanics and follow this with calibration demonstrations using these representations.
These calibration exercises illustrate the utility of the method in discovering the best-fit symmetry groups through sparse regression.
For these demonstrations we use neural network models \cite{fuhg2022learning} but the methods are equally adaptable to any suitable function representation.
The appendices provide numerical values for some of the characteristic tensor results.
We conclude with a summary of the findings and avenues for future work in \sref{sec:conclusion}.

\section{Related work} \label{sec:related}

\paragraph{Invariant theory}
Invariants and tensors have widespread use in mathematical physics, from quantum to continuum mechanics to general relativity. Typically, physical principles place restrictions on the form of these tensors. In this work, we compute the basis for tensor spaces subject to linear constraints arising from symmetry considerations. Previous works on this topic required careful analysis to obtain these bases \cite{karafillis1993general,itin2018irreducible}. Compared to these works, we propose simple, fully automated algorithms to obtain these tensor bases utilizing only group generators or Reynolds operators. Additionally, we leverage tools from ring theory to accelerate our algorithms. A related line of research focuses on computing the integrity basis for invariant functions of tensors \cite{olive2014isotropic,desmorat2021minimal,aguiar2023construction} or computing minimal generating sets for invariant rings \cite{kemper1999,sturmfels2008algorithms,king2013ring,derksen2015computational}.

\paragraph{Continuum mechanics}
Interest in constructing functions representing material symmetries arose in the 1950s concurrently with the inception of rational mechanics.
Rivlin \cite{rivlin1955further,smith1957anisotropic,smith1957stress,spencer1958theory,spencer1958finite,spencer1962isotropic,smith1963integrity,smith1964integrity,rivlin1969orthogonal},
Smith \cite{smith1957anisotropic,smith1957stress,spencer1962isotropic,smith1964integrity,kiral1974constitutive,smith1996irreducible},
Spencer  \cite{spencer1958finite,spencer1958theory,spencer1959further,spencer1959further,spencer1965isotropic,spencer1970note,spencer1987isotropic},
Boehler \cite{boehler1979simple,boehler1987representations,boehler1987rational,zheng1994description},
and
Wang \cite{wang1969general,wang1970new}
all made significant contributions.
After Rivlin and Ericksen \cite{rivlin1955stress} and others, investigations of complete representations for tensor polynomials were given attention and resulted in finite, complete expansions with the help of the Cayley-Hamilton theorem.
Typically, these representations take the form of coefficient functions dependent on scalar invariants of the inputs which multiply a tensor-valued basis which spans the output space.
The Wineman-Pipkin theorem \cite{wineman1964material} is a foundation of these representations and proves that tensor polynomials are complete representations of tensor functions.

This body of work introduced a number of concepts from group theory and related mathematics to the mechanics community.
\emph{Integrity} bases
\cite{boehler1987representations}
are sets of polynomial scalar invariants that can be used as complete inputs to scalar-valued functions such as the coefficient functions.
An integrity basis is \emph{irreducible} if no subset is a complete representation by itself \cite{jemiolo1991irreducible,smith1996irreducible}; however, polynomial relations, called \emph{syzgies}, may exist between the elements.
Irreducible bases are not unique, but provide some assurance of solvability and simplicity.
A \emph{functional} basis \cite{boehler1987representations}  is analogous to an integrity basis but allows non-polynomial invariants and hence can be more compact.
Tensor basis expansion of a tensor-valued function is irreducible if the integrity basis is irreducible and its elements are linearly independent.

The concept of a tensor characteristic of a symmetry group (a \emph{structural} tensor) was, to the authors knowledge, introduced by Smith and Rivlin \cite{smith1957anisotropic}.
Boehler \cite{boehler1979simple} extended the concept and
Zhang and Rychlewski provided the \emph{isotropization} theorem \cite{zhang1990structural}, which is the basis for using structural tensors in anisotropic response functions.
The theorem states that an anisotropic response function of a given input can be represented by an isotropic function of the same input and the appropriate structural tensor.
This allows the use of the wealth of fairly general isotropic tensor function representations, e.g. by Wang \cite{wang1969general,wang1970new} and Smith \cite{smith1970fundamental,smith1971isotropic,kiral1974constitutive}.
Prior to these developments, anisotropic function representations were designed directly (without the aid of the structural tensor input), which is a considerably more difficult and bespoke task.
Further perspective can be found in the collection by Boehler \cite{boehler1987introduction} and the most comprehensive collection of results by Zheng \cite{zheng1993representations,zheng1994theory}.

The Pipkin-Rivlin procedure \cite{pipkin1959formulation}, which relies on some intuitive reasoning, is apparently the basis for many of the existing results; however, some algorithmic methods have been proposed in limited contexts, for example Refs. \cite{xu1987computer,miehe1998comparison,itin2018irreducible}.

\section{Method} \label{sec:method}

First, we present two related methods, \Aref{alg:alg} and \Aref{alg:alg2}, for constructing tensors that satisfy linear constraints based on basic linear algebra. Specifically, we are interested in finding a basis for a tensor of a particular order subject to symmetry, trace, and transposition constraints.
Then we illustrate the use of \Aref{alg:alg} in a few examples and apply the method to the general problem of finding characteristic structure tensors and integrity bases.

\subsection{Preliminaries} \label{sec:pre}

Here, we introduce selected concepts from representation theory, ring theory, and mechanics.
In the following, we will denote tensors as $\Ab$, spaces as $\AA$, groups as $\As$, and algebras as $A$.

Since we focus on applications in mechanics, we work with $d$ dimensional real vector spaces, $\RE^d$.
Although the methods we develop are more general, we restrict ourselves to matrix groups, i.e., subgroups of the general linear group, $GL(d,\RE)$.
Within these matrix groups, we will work exclusively with subgroups of the orthogonal and special orthogonal groups because they describe the material symmetries encountered in mechanics \cite{zheng1994theory}.

The groups under consideration act on vectors in $\RE^d$ via matrix-vector multiplication, but they also act on vector spaces built from $\RE^d$.
In these developments, $n$-th order tensors are elements of the $n$-th fold tensor product of $\RE^d$ with itself, $\bigotimes_n\RE^d \equiv T^n(\RE^d)$.
Given a group, $\grp$, the group element $\Gb\in\grp$, acts on a tensor, $\Ab \in T^n(\RE^d)$, via the Kronecker product / group action, $\Gb \boxtimes \Ab$ (see \eref{eq:tp_action} for the formal definition).
We can also construct subspaces of $T^n(\RE^d)$ by projecting to spaces of totally symmetric, $S^n(\RE^d)$, or skew-symmetric tensors, $A^n(\RE^d)$. We will refer to the binary operations that produce these symmetric and skew-symmetric spaces as, $\otimes_s$ and $\otimes_a$, respectively.

The binary operations, $\otimes$, $\otimes_s$, and $\otimes_a$ produce the tensor $T(\RE^d)$, the symmetric tensor $S(\RE^d)$, and  the alternating algebras $A(\RE^d)$, respectively. These algebras are graded rings, i.e., they are rings that can be decomposed via the direct sum into vector spaces of homogeneous-order components. For example, the tensor algebra is composed of the tensor product spaces, $T(\RE^d) = {\bigoplus}_{n=0}^{\infty} \bigotimes_{i=0}^{n}\RE^d = T^0(\RE^d) \oplus T^1(\RE^d) \oplus \hdots$.  Finally, the Hilbert series is a formal power series for a ring whose coefficients measure the dimension of the graded pieces of the ring. For example, the Hilbert series of the tensor algebra in $d=3$ is,
\begin{equation}
\begin{aligned}
H(T(\RE^3),t) = \frac{1}{(1-3t)} = 1+t+3t+9t^2+27t^3+81t^4+\hdots
\end{aligned}
\end{equation}
and the coefficient of the $n$-th power of $t$ give the dimension of the space of $n$-th order tensors. See \cref{cox1997ideals} for more comprehensive details on ring theory.

\subsection{Methods for obtaining a basis for tensors of a given order subject to constraints} \label{sec:null_space}

We consider two complementary approaches for finding the basis for a vector space subject to constraints. In the first approach, we seek a basis for the nullspace of the linear operator subject to the constraints. In the second approach, we seek a basis for the image of a projection operator that satisfies the constraints.

Let $\AA \subseteq \bigotimes_n \RE^d$ be the space of $n$-th order tensors in $d$-space subject to $m$ linear constraints associated with the operator $L$, such that
\begin{equation} \label{eq:linear_op}
\begin{aligned}
&L: \bigotimes_n \RE^d \rightarrow \RE^m \\
&\ker L = \AA
\end{aligned}
\end{equation}
We can construct a basis for $\AA$
\begin{equation} \label{eq:rep}
\Ab = \sum_i c_i \Ab_i \ \text{with} \ c_i \in \RE \quad \forall \Ab \in \AA
\end{equation}
by finding the  the nullspace of $L$ ( the \emph{kernel} $\ker L$):
\begin{equation} \label{eq:null_space}
L[\Ab] = \mathbf{0} \quad \forall \Ab \in \AA
\end{equation}

Alternatively, if we have a projection operator $P$,
\begin{equation}\label{eq:proj}
\begin{aligned}
&P:\bigotimes_n\RE^d \rightarrow \bigotimes_n\RE^d \\
&\mathrm{im} P = \AA
\end{aligned}
\end{equation}
then we can construct a basis for $\AA = \mathrm{im} P$, where $\mathrm{im} P$ is the \emph{image} of $P$.
Note that both of the approaches can be used, i.e., if we have a projection operator and constraints such that $\AA = \mathrm{im} P \cap \ker L$, then we can sequentially compute, $L\Ab = \mathbf{0}; \forall \Ab \in \ker P$, or vice versa.

Being linear in $\Ab$, the action of the operator $L$ or $P$ on $\Ab$ can be converted to a matrix-vector form amenable to standard numerical algebra algorithms.
The vectorization operator $\eta$ provides an isomorphism that transforms the tensor $\Ab$ to a flat vector $\As \in \RE^{d^n}$,
\begin{equation} \label{eq:map}
\eta: \bigotimes_n \RE^d \to \RE^{d^n}.
\end{equation}
such as $\eta: \eb_i \otimes \eb_j \ldots \mapsto \eb'_{i+dj+ \ldots}$.
This allows us to rewrite \eref{eq:linear_op} as:
\begin{equation} \label{eq:eval_prob}
\begin{aligned}
&\mathsf{L}: \RE^{d^n} \rightarrow \RE^m\\
&\mathsf{L}[\As] = L \eta^{-1} [\As] = \mathsf{0} \quad \forall \mathsf{A} \in \hat{\AA} = \left\{ \eta \Ab \ | \ \Ab \in \AA \right\}.
\end{aligned}
\end{equation}
so that $\hat{\AA}$ is the nullspace of the operator $\mathsf{L}$.
We can obtain a basis for $\hat{\AA}$, $\left\{ \mathsf{A}_i \right\}$, and transform it to a basis for $\AA$ via the inverse of the vectorization operator, $\left\{ \Ab_i \right\} = \left\{ \eta^{-1} \mathsf{A}_i \right\} $. The transformation for the projection operator, $P$, follows analogously.

The algorithm summarized in \Aref{alg:alg} utilizes the singular value decomposition (SVD) \cite{golub2013matrix} to compute the nullspace of $L$. Note, as described, the method is restricted to homogeneous constraints $L[\Ab] = \mathbf{0}$.
An analogous, naive algorithm for computing the image of $P$ would be to simply compute the action of $P$ over a standard basis. However, in the next section, we will provide a more efficient randomized algorithm along with a refinement for \Aref{alg:alg}.

\begin{algorithm}[ht]
\textbf{Input}:
Linear constraint operator $L$ and a bijective map $\eta $ for a tensor of order $n$ \\
\begin{enumerate}
\item
Translate the problem using map $\eta$: construct a matrix, $\Ms \in \RE^{m \times d^n}$, by evaluating $\mathsf{L} = L \circ \eta^{-1}$ on the standard basis, $M_{.j} = L\eta^{-1}(\eb_j)$. \\
\item
Next, find a basis for nullspace of $\Ms$ by constructing its SVD, $\Ms = \Us \Ds \Vs^T$, and taking as basis vectors, $\vs_j$, the rows of $\Vs^T$ associated with singular values of 0 (i.e. those with singular values less than tolerance determined by a gap in the singular values). \\
\item
Finally, construct the basis for $\Ab$ by applying the isomorphism, $\Ab_j = \eta^{-1} \vs_j$.  \\
\end{enumerate}
\textbf{Result:}
A basis $\{ \Ab_i \}$ for solutions of $L[\Ab] = \mathbf{0}$.
\caption{Construction of linear constrained tensors, solutions of \eref{eq:linear_op}}
\label{alg:alg}
\end{algorithm}

\subsection{Predicting the dimension of the nullspace}

We can improve on \Aref{alg:alg} by using a truncated SVD to obtain the nullspace of $L$ if we have an upper bound for the dimension of the nullspace. Given a group, the elements in a ring, $R$, that are invariant to all group elements form the subring, $R^\grp$.
Molien's formula \cite{sloane1977error, sturmfels2008algorithms,drensky2023invariant} for the full tensor algebra, the symmetric algebra, and alternating algebra give the Hilbert series for the invariant subrings:
\begin{eqnarray} \label{eq:moliens}
H(T(\RE^d)^\grp,t) &=& \frac{1}{|\grp|}\sum_{g\in\grp} (1-t\,\mathrm{tr}(g))^{-1}\\
H(S(\RE^d)^\grp,t) &=& \frac{1}{|\grp|}\sum_{g\in\grp} (\det(I-tg))^{-1}\\
H(A(\RE^d)^\grp,t) &=& \frac{1}{|\grp|}\sum_{g\in\grp} (\det(I+tg))
\end{eqnarray}
respectively. 
Generally, the invariant vector spaces belong to $T(\RE^d)^\grp$, so their dimensions are bounded by $H(T(\RE^d)^\grp,t)$.
We can tighten the bounds if we are looking for spaces in $S(\RE^d)^\grp$ or $A(\RE^d)^\grp$ by using their respective Hilbert series.

For more general tensors subject to transposition symmetries, we can also obtain tighter bounds if we restrict ourselves to a given tensor order. The Reynolds operator is
\begin{equation}
\begin{aligned}
\mathrm{Rey}: \RE^d \rightarrow \RE^d \quad \\
\mathrm{Rey} = \frac{1}{|\grp|}\sum_{\Gb\in\grp} \Gb
\end{aligned}
\end{equation}
where $\mathrm{Rey}$ is a projection onto the invariants in $\RE^d$.
For the compact groups, equivalent Reynolds operators can be defined using Haar measures \cite{ecker2024haar}.
Therefore, the sum of its eigenvalues $\{ \lambda_i \}$ count the dimension of the nullspace for invariant vectors, $\tr\ \mathrm{Rey} = \sum_i \lambda_i = \dim \mathrm{im\ Rey} = \dim \ker L$.
As discussed in Section~\ref{sec:pre}, the action of group elements can be extended to tensor, symmetric, and alternating powers.
The eigenvalues of the Reynolds operators on these spaces can be calculated as well.
Suppose the group element, $\Gb$, has eigenvalues and eigenvectors, $\lambda_i^{\Gb}$ and $\omega_i^{\Gb}$, and we have a vector space, $\VV$, built from elements of $\WV$ with the multiplication operator, $\circ$.
The products, $\lambda_i^\Gb \lambda_j^\Gb$ and $\omega_i^\Gb \circ \omega_j^\Gb$ are then eigenvalues and eigenvectors of the action of $\Gb$ on elements of $\VV$ because $\Gb \boxtimes (\omega_i^\Gb \circ \omega_j^\Gb) = \Gb\omega_i^\Gb \circ \Gb\omega_j^\Gb = \lambda_i^\Gb\lambda_j^\Gb \omega_i^\Gb \circ \omega_j^\Gb$.
The set of unique products, $\omega_i^\Gb \circ \omega_j^\Gb$, form a basis for $\VV$. Because $\Gb$ is orthogonal, so is its action in $\VV$, so the action is also diagonalizable and we can compute the trace of the action by summing its eigenvalues.
We can compute the dimension of the image of the Reynolds operator on $\VV$ and therefore the invariant subspace, $\VV^\grp$, as,
\begin{equation}
\begin{aligned}
\mathrm{Rey}^\VV: \VV &\rightarrow \VV  \quad \\
\dim \VV^\grp = \dim \mathrm{im} \mathrm{Rey}^\VV = \tr\ \mathrm{Rey}^\VV &= \frac{1}{|\grp|} \sum_{\Gb \in \grp} \tr \Gb \boxtimes
= \frac{1}{|\grp|} \sum_{\Gb \in \grp} \quad \sum_{i,j \forall \omega^\Gb_i \circ \omega^\Gb_j} \lambda_i^\Gb \lambda_j^\Gb
\end{aligned}
\end{equation}

\paragraph{Example}
\textit{The elastic moduli have Cauchy transpositional symmetries.
Karafillis and Boyce \cite{karafillis1993general} sought tensors with similar transposition symmetries for anisotropic yield functions, $\Lbb_{ijkl}=\Lbb_{jikl}=\Lbb_{ijlk}=\Lbb_{klij}$ (see \sref{sec:illustrations} for more details).
This space is formed by the twice symmetric square of $\RE^d$, $S^2(S^2(\RE^3))$. We can form this space by taking the tensor product of $\RE^d$ with itself and symmetrizing to produce $S^2(\RE^3)$ and then taking the tensor product of $S^2(\RE^3)$ with itself and symmetrizing.
If we obtain the eigenvalues of group elements, $\lambda^\Gb_i$, we can compute the dimension with:
\begin{equation}
\begin{aligned}
\dim S^2(S^2(\RE^3))^\grp = \frac{1}{|\grp|} \sum_{\Gb \in \grp} \quad \sum_{k\leq l} \mu_k^\Gb \mu_l^\Gb,
\\ \mu^\Gb_k = \lambda_i^\Gb \lambda_j^\Gb \quad \mathrm{for}\ i \leq j.
\end{aligned}
\end{equation}
}

Given these bounds for the dimension of the invariant subspace, we can modify \Aref{alg:alg} to use a truncated SVD instead of the full SVD. For high dimensional and high order tensors, the truncated SVD is substantially more computationally efficient, as we demonstrate in \sref{sec:timings}. Our implementation utilizes the LOBPCG eigensolver \cite{knyazev2001toward} to find the nullspace.

\subsection{A randomized algorithm} \label{sec:fast} 
Alternatively, we can seek the image of the Reynolds operator to find an invariant subspace. We propose the matrix-free randomized shown in  \Aref{alg:alg2}. This algorithm is a matrix-free adaptation of the randomized range finder algorithm utilizing a sparse test matrix, see \cite{halko2011finding}. In practice, we find that $m\approx \log \dim V$ and $p\approx 5$ provide sufficient hyperparameters. Note that this algorithm is trivially parallelizable \cite{halko2011finding}.

\begin{algorithm}[ht]
\textbf{Input}:
Group $\grp$, the vector space, $\VV$, number of samples $m$, rank $r$, and oversampling parameter $p$\\
\begin{enumerate}
\item Initialize $A = \{\}$
\item Randomly select $m$ basis elements from $\VV$, $\mathbf{b}_m$
\item Rotate the basis elements by every element, $\Gb \in \grp$ and add $\sum_m \Gb \boxtimes \mathbf{b}_m$ to the set $A$
\item Repeat steps 2-4 until $|A| = r+p$
\item Compute a rank-revealing QR decomposition of $A$ and retain the columns of $Q$, $\{\Ab_i\} = \{Q_i\}$
\end{enumerate}
\textbf{Result:}
A basis $\{ \Ab_i \}$ for the image of $\mathrm{Rey}^\VV$.
\caption{Construction of projected tensors, solutions of \eref{eq:proj}}
\label{alg:alg2}
\end{algorithm}

\subsection{Timings}\label{sec:timings}

In this section, we compare the timings of Alg.\ref{alg:alg} using the full SVD, \Aref{alg:alg} using the truncated SVD, and \Aref{alg:alg2}.

We find the invariant subspace with respect to the cubic group of the symmetric power of the tensor square for varying symmetric powers up to the ninth power. Our results are shown in \tref{tab:timing}. These were performed on a 32GB A100 GPU in Jax \cite{jax2018github}. Although the full and truncated SVD approaches are faster, the randomized method is more memory efficient. Additionally, the randomized method is more easily parallelized.
\begin{table}[h]
\centering
\begin{tabular}{c|c|c|c|c|c|c|c}
Order & 3 & 4 & 5 & 6 & 7 & 8 & 9 \\
\hline
Full SVD & 0.438s & 1.57s & OOM & OOM & OOM & OOM & OOM \\
Truncated SVD & 0.037s & 0.084s & 0.584s & OOM & OOM & OOM & OOM \\
Randomized Range finder & 7.59s & 26.7s & 74.4s & 268s &  584s & 1636s & 6252s \\
\end{tabular}
\caption{Timings for three algorithms. OOM refers to out-of-memory.}
\label{tab:timing}
\end{table}

\section{Application to mechanics} \label{sec:mechanics}
Now we apply the method to common problems in mechanics.
First, we introduce some notation and identities to simplify the following developments.
Then we illustrate the basics of the method by finding elastic modulus and stress transformation tensors, see e.g. \cref{karafillis1993general}, that satisfy the common crystal symmetries.
Next, we define structural tensors formally and then describe how the method can be used to find all invariant tensors for any group.

\subsection{Preliminaries}

\paragraph{Notation}
In the following, we adopt the notation:
$\Abb$ for a fourth order tensor, $\Ab$ for a second order tensor, $\ab$ for a vector (1st-order tensor), and $\As$ for a matrix of any order.

The Euler-Rodriguez formula:
\begin{equation} \label{eq:rodriguez}
\rotation_\ab(\theta)
= \Ib + \sin(\theta) \Ab + (1-\cos(\theta)) \Ab^2
\end{equation}
provides the rotation of $\theta$ about $\ab$,
where $\Ib$ is the second order identity tensor, $\Ab = \perm \ab$ and $\varepsilonb$ is the 3rd-order permutation tensor.
Also the short-hand
\begin{equation}
\rotation_i(\theta)
\equiv \rotation_{\eb_i}(\theta)
\end{equation}
will be used for rotations about the Cartesian axes,
\begin{equation}
\reflection_\ab = \Ib - 2 \ab \otimes \ab
\end{equation}
for general reflections, and
\begin{equation}
\reflection_i = \reflection_{\eb_i}
\end{equation}
for reflections through the Cartesian axes.
Lastly, we denote two-dimensional identity tensors/projectors as
\begin{equation}
\Ib_\ab = \Ib - \ab \otimes \ab
\end{equation}
and
\begin{equation}
\Ib_i = \Ib_{\eb_i} \ .
\end{equation}

The fourth order tensors that are invariant to the isotropic group (the full \SOthree\ or \Othree\ group) are well-known:

\begin{equation}\label{eq:IJK}
\begin{aligned}
\Ibb &= \Ib \otimes \Ib =  \delta_{ij} \delta_{kl} \eb_i \otimes \eb_j \otimes \eb_k \otimes \eb_l\\
\Jbb &= ( \delta_{ik} \delta_{jl} +  \delta_{il} \delta_{jk} ) \eb_i \otimes \eb_j \otimes \eb_k \otimes \eb_l  \\
\Kbb &=  ( \delta_{ik} \delta_{jl} -  \delta_{il} \delta_{jk} ) \eb_i \otimes \eb_j \otimes \eb_k \otimes \eb_l .
\end{aligned}
\end{equation}
For certain other symmetries, Zheng \cite{zheng1993representations} provides well-established fourth order structural tensors.
The cubic group is characterized by:
\begin{equation}
\label{eq:O}
\Obb_h \equiv
\sum_{i=1}^3 \eb_i \otimes \eb_i \otimes \eb_i \otimes \eb_i
= \delta_{ij} \delta_{jk} \delta_{kl} \delta_{li}  \eb_i \otimes \eb_j \otimes \eb_k \otimes \eb_l \ .
\end{equation}
Likewise, the tetragonal group has a structural tensor:
\begin{equation} 
\label{eq:D}
\Dbb_{4h} = \sum_{i,j,k,l=1}^2 \operatorname{Re}(\imath^{(i+j+k+l)}) \eb_i \otimes \eb_j \otimes \eb_k \otimes \eb_l  \ .
\end{equation}

\paragraph{Identities}
A number of identities are apparent
\begin{eqnarray}
\reflection_i^{2n} &=& \Ib \\
\rotation_i(\pi)^{2n} &=& \Ib \\
\rotation_i(\pi/n)^{n} &=& \Ib \\
\rotation_1(\pi) \rotation_2(\pi) \rotation_3(\pi) &=& \Ib \\
\reflection_2(\pi) \reflection_1(\pi) \reflection_3(\pi) &=& -\Ib
\end{eqnarray}
and facilitate understanding the set of generators that characterize subgroups of \Othree, as will be demonstrated in \sref{sec:crystal_groups}.

\subsection{Illustrations} \label{sec:illustrations}

Many constitutive properties, such as elastic modulus tensors, take the form of a high-order tensor subject to linear constraints.
These typically result from the linearization of some physical response function with respect to its inputs, e.g., the elastic (tangent) modulus $\Cbb$
\begin{equation}
\Cbb = \partialb_\Eb \Sb
\end{equation}
is the linearization of a stress $\Sb$ with respect to a chosen strain measure $\Eb$.

\paragraph{Elastic modulus tensor}
We can use the proposed method to construct a general elastic modulus tensor $\Cbb$ that has the Cauchy symmetries
\begin{eqnarray}
\Cbb_{ijkl} &=& \Cbb_{jikl} \\
\Cbb_{ijkl} &=& \Cbb_{ijlk} \\
\Cbb_{ijkl} &=& \Cbb_{klij}
\end{eqnarray}
which result from generalized transpose operators,
and those of the cubic symmetry group
\begin{eqnarray}
\rotation_1(\pi/2)_{ai} \rotation_1(\pi/2)_{bj}  \rotation_1(\pi/2)_{ck}  \rotation_1(\pi/2)_{dl} \Cbb_{ijkl} &=& \Cbb_{abcd} \\
\rotation_2(\pi/2)_{ai} \rotation_2(\pi/2)_{bj}  \rotation_2(\pi/2)_{ck}  \rotation_2(\pi/2)_{dl} \Cbb_{ijkl} &=& \Cbb_{abcd} \\
\rotation_3(\pi/2)_{ai} \rotation_3(\pi/2)_{bj}  \rotation_3(\pi/2)_{ck}  \rotation_3(\pi/2)_{dl} \Cbb_{ijkl} &=& \Cbb_{abcd}
\end{eqnarray}
which result from generators $\grp = \{ \rotation_i(\pi/2), i=1,2,3\}$.

These constraints, which are linear in $\Cbb$ and homogeneous, result in
\begin{equation}
\Cbb = \sum_\alpha c_\alpha \Cbb_\alpha
\end{equation}
where $c_\alpha$ are coefficients specific to the application.

Our algorithm finds $\Cbb_\alpha$ as
\begin{eqnarray}
\Cbb_1 &=&   5.9023957 \, \Ibb -3.1952064 \, \Jbb \\
\Cbb_2 &=&  5.1072836 \, \Ibb + 4.9463570 \, \Jbb \\
\Cbb_3 &=& \Obb_h
\end{eqnarray}
which are expressible in the more familiar $\Ibb$, $\Jbb$, and $\Obb_h$, which is equivalent to:
\begin{equation} \label{eq:cubic_result}
\begin{aligned}
\Cbb
&=( 5.9023957 c_{1} +  5.1072836 c_{2})  \Ibb + (-3.1952064 c_{1} + 4.9463570 c_{2}  ) \Jbb + c_{3} \Obb_h \, .
\end{aligned}
\end{equation}
We can compare this to the general familiar representation of cubic symmetries \cite{landau2012theory} with the three independent elastic constants $\mu$, $\lambda$, and $\alpha$
\begin{equation} \label{eq:cubic_std}
\Cbb = \mu \Ibb + \lambda \Jbb + \alpha \Obb_h
\end{equation}
and can match terms or simply reparameterize the result \eqref{eq:cubic_result} as \eref{eq:cubic_std}.

\paragraph{Stress transformation tensor}

Plasticity models require a yield surface to circumscribe the elastic response region:
\begin{equation} \label{eq:yield}
f( \Sb) \le \Upsilon
\end{equation}
To imbue a yield surface with material anisotropy some authors \cite{hill1950mathematical,barlat1991six,karafillis1993general} transform the stress $\Sb$
\begin{equation}
\Sb' = \Lbb \Sb
\end{equation}
before applying the yield function $f$ in \eref{eq:yield}.
This implicitly forms invariants between $\Lbb$ and $\Sb$.

As in Karafilis and Boyce \cite{karafillis1993general}, we can find the components of a fourth order tensor $\Lbb$ subject to,
\begin{eqnarray}
\Lbb_{ijkl} &=& \Lbb_{ijlk} = \Lbb_{jikl} \\
\Lbb_{ijkl} &=& \Lbb_{klij}\\
\Lbb_{ijkk} &=& 0,
\end{eqnarray}
and various crystal symmetries.

We demonstrate the truncated SVD approach here and first compute the dimension of $S^2(S^2(\RE^3))^\grp$ as discussed in \sref{sec:fast}.
The summary of results shown in \tref{tab:karafillis} indicate we find identical results with the exception of the monoclinic class where we find one more basis element than \cref{karafillis1993general}. Since our bases are orthogonal they are independent.
Note that the number of components is less than the dimension of $S^2(S^2(\RE^3))^\grp$ because the additional traceless conditions further restrict the subspace.

\begin{table}[h]
\centering
\begin{tabular}{|c|c|c|}
\hline
Group & Components & $\dim S^2(S^2(\RE^3))^\grp$\\
\hline
Triclinic & 15 & 21\\
Monoclinic & 9 & 13\\
Orthotropic & 6 & 9\\
Trigonal & 4 & 6\\
Tetragonal & 4 & 6\\
Transverse isotropic & 3 & 5\\
Cubic & 2 & 3\\
Isotropic & 1  & 2\\
\hline
\end{tabular}
\caption{Summary of bases found for an anisotropic yield function using the definitions in \cref{karafillis1993general}. }
\label{tab:karafillis}
\end{table}

\paragraph{Example involving antisymmetric components}

For our final example, we seek the basis for sixth order invariant tensors in $A^2(S^3(\RE^3))$. These tensors have the following symmetries,
\begin{eqnarray}
\Abb_{ijklmn} &=& \Abb_{jiklmn} = \Abb_{ikjlmn} = \Abb_{ijkmln} = \Abb_{ijklnm} \\
\Abb_{ijklmn} &=& -\Abb_{lmnijk}
\end{eqnarray}
We reuse the groups examined for the stress transformation tensor and obtain bases with the dimensions listed in Table~\ref{tab:manufactured} using the truncated SVD approach. Note that in this example, the bound, $\dim A^2(S^3(\RE^3))^\grp$ is tight because we do not have any trace constraints.

\begin{table}[h]
\centering
\begin{tabular}{|c|c|}
\hline
Group & Components and $\dim A^2(S^3(\RE^3))^\grp$\\
\hline
Triclinic & 45\\
Monoclinic & 21\\
Orthotropic & 9\\
Trigonal & 6\\
Tetragonal & 4\\
Transverse isotropic & 2\\
Cubic & 1\\
Isotropic & 0 \\
\hline
\end{tabular}
\caption{Summary of bases.}
\label{tab:manufactured}
\end{table}

\subsection{Structural tensors as tensors characteristic of symmetry groups} \label{sec:structural_tensors}
We now proceed to apply the method to find \emph{structural} tensors for a given symmetry group, but first we introduce some notation and identities to simplify the following developments.

Let $\grp$ be a matrix group acting on an $\RE^d$ vector space.
The action of a group element on an $n$-th order tensor $\Bb$ is obtained via contracting it with an $n$-th fold tensor product of the group element,
\begin{equation}\label{eq:tp_action}
\generator \boxtimes \Bb =  B_{ijk\ldots} \generator \eb_i \otimes \generator \eb_j \otimes \generator \eb_k \otimes \ldots
\end{equation}
as in the Kronecker product.
By definition, a structural tensor $\Ab$  is left invariant under group action
\begin{equation} \label{eq:structural_tensor}
\generator \boxtimes \Ab  = \Ab
\  \forall \ \generator \in \grp
\end{equation}
Note, (a) any scaling of $\Ab$ is also a structural tensor
\begin{equation}
\generator \boxtimes c \Ab
= c\generator \boxtimes \Ab
= c \Ab
\  \forall \ \generator \in \grp \ ,
\end{equation}
hence (b) a tensor in the span of a set of structural tensors is a structural tensor:
\begin{equation}
\generator \boxtimes \sum_i c_i \Ab_i
= \sum_i c_i \generator \boxtimes \Ab_i
= \sum_i c_i  \Ab_i
\  \forall \ \generator \in \grp
\end{equation}
A structural tensor also needs to exclude members of the complement (in the set theoretic sense) of the group $\grp$ i.e.
\begin{equation} \label{eq:exclude_groups}
[ \generator \boxtimes  - \Ib \boxtimes ] \Ab  \neq \mathbf{0}  \quad \forall \ \generator \in \text{\SOthree}\setminus \grp
\end{equation}
We will call a tensor that is both \emph{invariant} \eref{eq:structural_tensor} and \emph{exclusive} \eref{eq:exclude_groups} a \emph{characteristic} tensor of group $\grp$, or more commonly a \emph{structural} tensor.

A simple example of a structural tensor is the axial vector $\ab$ for the group of all rotations around this axis  $\grp = \{ \rotation_{\ab}(\theta) \}$.
These rotations leave the axial vector $\ab$ unchanged:
\begin{equation}
\rotation_{\ab}(\theta) \ab = \ab
\end{equation}
i.e. $\ab$ is the unit eigenvector of all $\{ \rotation_{\ab}(\theta) \}$.
Hence, this axial vector is a solution of \eref{eq:structural_tensor} and a structural tensor for this group.
Note $\Ab \equiv \ab$ excludes the inverse identity $-\Ib$, while $\Ab = \ab\otimes\ab$ includes it.

\subsection{Application to finding tensors that are characteristic of symmetry groups} \label{sec:structural_tensors_method}
Using the results of \sref{sec:null_space}, in particular \Aref{alg:alg},
the problem of finding a structural tensor or set of structural tensors that satisfy \eref{eq:structural_tensor} can be reframed as
\begin{equation} \label{eq:struct_operator}
\mathsf{L}[\mathsf{A}] = \begin{bmatrix}
\eta \circ [ \generator_1 \boxtimes - \Ib \boxtimes] \circ \eta^{-1}\\
\eta \circ [ \generator_2 \boxtimes - \Ib \boxtimes] \circ \eta^{-1}\\
\vdots \\
\eta \circ [ \generator_p \boxtimes - \Ib \boxtimes] \circ \eta^{-1}\\
\end{bmatrix} \mathsf{A} =  \Gs \As =  \mathsf{0}
\end{equation}
from the constraints
\begin{equation}\label{eq:struct_constraints}
[ \generator \boxtimes - \Ib \boxtimes] \Ab  = \mathbf{0}  \quad \forall \ \generator \in \grp
\end{equation}
where $\Ib$ is the identity element of $\grp$  and the constraints associated with the multiple elements of $\grp$ have been stacked in $\Gs$.
This linear system has previously been used to compute invariant subspaces as part of algorithms for computing integrity bases \cite{derksen2015computational}.
The constraints can be reduced to a finite set of generators, even for reductive Lie groups for which a finite set of topological generators exist; hence,  we rely on well-known sets of group generators to form the constraints.

Typically the procedure produces multiple bases for the nullspace and hence any tensor in the span of these tensors is a valid structural tensor if it is also exclusive, or, as we show in \sref{sec:mechanics} the entire basis (or a sub-basis) can be used in the representation of a physical response function.
Furthermore, to obtain a single representative tensor, we can simplify (in the sense of minimizing non-zero components) a single tensor in the span of the basis by sparse optimization via \Aref{alg:simplify}.
\begin{algorithm}[h!]
\textbf{Input:} a $n$-th order basis for a symmetry group $\grp$
\begin{enumerate}
\item[] Optimize objective
\begin{equation*}
c_a = \operatorname{argmin}_{c_a} \| \Bb \|_{1}  \quad \text{s.t.} \quad
\| \Bb \|_{\infty} = 1
\end{equation*}
to obtain the coefficients $c_a$,
where
\begin{equation*}
\Bb = \sum_a c_a \Bb_a
\end{equation*}
is in the span of the basis $\{\Bb_a\}$:
\end{enumerate}
\textbf{Result}: a characteristic tensor $\Bb$ with the fewest non-zero components.
\caption{Simplication of a single characteristic tensor via sparsification}
\label{alg:simplify}
\end{algorithm}

To verify that the basis for a subgroup is contained in another, we can check that every basis element of the smaller group is in the span of the larger group:
\begin{equation} \label{eq:inspan}
\Abb = \sum_a c_a \Abb_a  \quad | \quad
\cs = [ \As^T \As ]^{-1} \As^T \bs
\end{equation}
where $\cs\equiv [c_a]$, $\As = [\Abb_a]$, and $\bs\equiv [\Abb]$,  are vectorized versions of the coefficients, basis of the larger group, and basis element of the smaller group, respectively.
Or, equivalently, we can check that the rank of the basis of the subgroup is equal to the rank of the union of the subgroup and group bases.

\subsection{The structure tensors for all basic crystal groups} \label{sec:crystal_groups}

In this demonstration, we use the methods of \sref{sec:structural_tensors_method} to determine all the characteristic tensors of order $n\le 4$ for the usual crystal symmetry groups (subgroups of \Othree).
\tref{tab:crystal_groups}a summarizes the results for the \SOthree\ group definitions given in Karafillis and Boyce \cite{karafillis1993general} Table 1.
Likewise \tref{tab:crystal_groups}b summarizes the \Othree\ results for Table 7 in Zheng\cite{zheng1993representations}.
For \tref{tab:crystal_groups}b we use the largest set of generators for class, e.g. for tetragonal we use Zheng's \emph{tetragonal 15} which includes inversions $-\Ib$.
Largely the two definitions differ by the addition of the inversions in our selection from Zheng; however, many groups differ in more substantial ways.
For instance, the cubic group in Zheng employs a $\rotation_\cb(2\pi/3)$ generator that cannot be generated from the Karafillis and Boyce generators in \tref{tab:crystal_groups}a.
The inclusion of $-\Ib$ substantially reduces the bases for the lower symmetry groups and has little or no effect on the higher symmetry groups.

\begin{figure}
\centering

\includegraphics[width=0.5\linewidth]{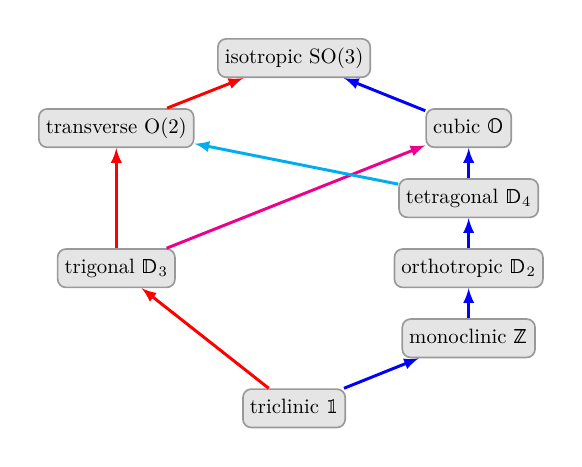}
\caption{Diagram of the poset of \SOthree\ subgroups (adapted from Auffray \etal \cite{auffray2011invariant}).
Note trigonal is a subgroup of cubic if the cube diagonal $\cb = (\eb_1+\eb_2+\eb_3)/\sqrt{3}$ is aligned with the rotation axis of the trigonal generator (usually $\eb_3$).}
\label{fig:SO3_poset}
\end{figure}

From the partially order set (\emph{poset}) diagram for \SOthree\ in \fref{fig:SO3_poset} and the generators in \tref{tab:crystal_groups}, the expected nesting/embedded on the subgroups is apparent.
In fact, we can use the poset to check the exclusivity condition \eqref{eq:exclude_groups}.
For example, the embedding triclinic $\subset$ monoclinic $\subset$ orthotropic $\subset$ tetragonal $\subset$ cubic $\subset$ isotropic is apparent from \fref{fig:SO3_poset}.
We verify that this partial ordering is preserved in the bases found by the proposed method.
Furthermore, we use Molien's formula \eqref{eq:moliens} to check that we find the correct number of basis elements.

\begin{table}[h]
\begin{subtable}[c]{0.9\textwidth}
\centering
\begin{tabular}{|c|c|}
\hline
Group & Generators \\
\hline
triclinic     &  $\Ib$ \\
monoclinic    & $\rotation_1(\pi)$  \\
orthotropic   & $\rotation_1(\pi),  \rotation_3(\pi)$  \\
trigonal      & $\rotation_3(2\pi/3)$ \\
tetragonal    & $\rotation_1(\pi), \rotation_3(\pi/2)$  \\
cubic         & $\rotation_1(\pi/2), \rotation_2(\pi/2), \rotation_3(\pi/2)$  \\
{\bf transverse isotropic} & $\rotation_1(\pi), \rotation_3(\theta)$ \\ 
{\bf isotropic}     & $\rotation_1(\theta_1), \rotation_2(\theta_2), \rotation_3(\theta_3)$ \\ 
\hline
\end{tabular}%
\begin{tabular}{|cccc|}
\hline
4 &  3&  2&  1\\
\hline
40/81  &  14/27  &   4/ 9  &   2/ 3 \\
33/41  &  10/13  &   2/ 5  &   0/ 1 \\
21/21  &   6/ 6  &   2/ 3  &   0/ 0 \\
10/11  &   0/ 3  &   0/ 2  &   0/ 0 \\
13/14  &   1/ 4  &   0/ 2  &   0/ 0 \\
4/ 4  &   0/ 1  &   0/ 1  &   0/ 0 \\
10/10  &   4/ 4  &   2/ 2  &   1/ 1 \\
3/ 3  &   0/ 0  &   1/ 1  &   0/ 0 \\
\hline
\end{tabular}
\caption{\SOthree}
\end{subtable}

\begin{subtable}[c]{0.9\textwidth}
\centering
\begin{tabular}{|c|c|}
\hline
Group & Generators \\
\hline
triclinic     &  $\Ib, -\Ib$ \\
monoclinic    & $\rotation_3(\pi), -\Ib$  \\
orthotropic   & $\reflection_1, \reflection_2, \rotation_3(\pi), -\Ib$  \\
trigonal      & $\reflection_2, \rotation_3(2\pi/3), -\Ib$ \\
tetragonal    & $\reflection_1, \reflection_2, \rotation_3(\pi/2), -\Ib$  \\
cubic         & $\rotation_\cb(2\pi/3), \rotation_1(\pi/2), \reflection_2, -\Ib$  \\
{\bf transverse isotropic} & $\rotation_1(\pi), \rotation_3(\theta)$ \\ 
{\bf isotropic}     & $\rotation_1(\theta_1), \rotation_2(\theta_2), \rotation_3(\theta_3)$ \\ 
\hline
\end{tabular}%
\begin{tabular}{|cccc|}
\hline
4 &  3&  2&  1\\
\hline
40/81  &   0/ 0  &   4/ 9  &   0/ 0 \\
32/41  &   0/ 0  &   2/ 5  &   0/ 0 \\
19/21  &   0/ 0  &   1/ 3  &   0/ 0 \\
12/14  &   0/ 0  &   0/ 2  &   0/ 0 \\
7/11  &   0/ 0  &   0/ 2  &   0/ 0 \\
4/ 4  &   0/ 0  &   0/ 1  &   0/ 0 \\
10/10  &   0/ 0  &   2/ 2  &   0/ 0 \\
3/ 3  &   0/ 0  &   1/ 1  &   0/ 0 \\
\hline
\end{tabular}
\caption{\Othree}
\end{subtable}

\caption{
(a) \SOthree\ symmetry groups from \cref{karafillis1993general}
and
(b) \Othree\ symmetry groups from Zheng \cite{zheng1993representations}.
and resulting structural tensors for orders up to 4.
Here $\cb \equiv \eb_1 + \eb_2 + \eb_3$.
Groups with non-finite generators are listed in {\bf bold}.
For the infinite groups, we use a single sample $\theta = \arccos(-3/5)$ which is not commensurate with $2\pi$.
The two numbers in the format A/B in the last 4 columns represent:
B is the number of invariant bases and A is the number of bases that exclude subgroups of \SOthree\ that are not subgroups of the specified group.
}
\label{tab:crystal_groups}
\end{table}

We focus on the two even-ordered basis sets ($n$=2,4):

\paragraph{Second order structure tensors}

We see in \tref{tab:2nd_order} that the structure tensors for \SOthree\ are largely composed of dyads with integer coefficients and many are recognizable, e.g., the second orthotropic entry is $-\rotation_1$ and the first transverse entry is $\Ib_3$.
It is also apparent that the nesting of groups is not maintained if the exclusive nature of a characteristic tensor is enforced, e.g. the cubic tensor is $-\Ib$.
The finding that second order tensors are inadequate to characterize all the crystal subgroups is corroborated by \tref{tab:2nd_order} for the \Othree\ Zheng generators (\tref{tab:crystal_groups}b).
In fact, second order tensors are only sufficient to characterize triclinic, monoclinic, orthotropic, transverse, and isotropic groups.

\begin{table}[]
\centering
\begin{subtable}[c]{0.9\textwidth}
\centering
\begin{tabular}{|c|cl|}
\hline
Group & count & tensors \\
\hline
\hline
triclinic & 4 & $ 1.0 \mathbf{e}_1 \otimes \mathbf{e}_3  $, \\
& & $ 1.0 \mathbf{e}_2 \otimes \mathbf{e}_3  $, \\
& & $ 1.0 \mathbf{e}_3 \otimes \mathbf{e}_1  $, \\
& & $ 1.0 \mathbf{e}_3 \otimes \mathbf{e}_2  $, \\
\hline
monoclinic & 2 & $ 1.0 \mathbf{e}_1 \otimes \mathbf{e}_2   + 1.0 \mathbf{e}_2 \otimes \mathbf{e}_1  $, \\
& & $ -1.0 \mathbf{e}_1 \otimes \mathbf{e}_2   + 1.0 \mathbf{e}_2 \otimes \mathbf{e}_1  $, \\
\hline
orthotropic & 2 & $ 1.0 \mathbf{e}_2 \otimes \mathbf{e}_2  -1.0 \mathbf{e}_3 \otimes \mathbf{e}_3  $, \\
& & $ 1.0 \mathbf{e}_1 \otimes \mathbf{e}_1  -0.5 \mathbf{e}_2 \otimes \mathbf{e}_2  -0.5 \mathbf{e}_3 \otimes \mathbf{e}_3  $, \\
\hline
transverse & 2 & $ 1.0 \mathbf{e}_1 \otimes \mathbf{e}_1   + 1.0 \mathbf{e}_2 \otimes \mathbf{e}_2  $, \\
& & $ 1.0 \mathbf{e}_3 \otimes \mathbf{e}_3  $, \\
\hline
isotropic & 1 & $ 1.0 \mathbf{e}_1 \otimes \mathbf{e}_1   + 1.0 \mathbf{e}_2 \otimes \mathbf{e}_2   + 1.0 \mathbf{e}_3 \otimes \mathbf{e}_3  $, \\
\hline
\end{tabular}
\caption{SO3}
\end{subtable}
\begin{subtable}[c]{0.9\textwidth}
\centering
\begin{tabular}{|c|cl|}
\hline
Group & count & tensors \\
\hline
\hline
triclinic & 4 & $ 1.0 \mathbf{e}_1 \otimes \mathbf{e}_3  $, \\
& & $ 1.0 \mathbf{e}_2 \otimes \mathbf{e}_3  $, \\
& & $ 1.0 \mathbf{e}_3 \otimes \mathbf{e}_1  $, \\
& & $ 1.0 \mathbf{e}_3 \otimes \mathbf{e}_2  $, \\
\hline
monoclinic & 2 & $ -1.0 \mathbf{e}_1 \otimes \mathbf{e}_2   + 1.0 \mathbf{e}_2 \otimes \mathbf{e}_1  $, \\
& & $ 1.0 \mathbf{e}_1 \otimes \mathbf{e}_2   + 1.0 \mathbf{e}_2 \otimes \mathbf{e}_1  $, \\
\hline
orthotropic & 1 & $ 1.0 \mathbf{e}_1 \otimes \mathbf{e}_1  -1.0 \mathbf{e}_2 \otimes \mathbf{e}_2  $, \\
\hline
transverse & 2 & $ 1.0 \mathbf{e}_1 \otimes \mathbf{e}_1   + 1.0 \mathbf{e}_2 \otimes \mathbf{e}_2  $, \\
& & $ 1.0 \mathbf{e}_3 \otimes \mathbf{e}_3  $, \\
\hline
isotropic & 1 & $ 1.0 \mathbf{e}_1 \otimes \mathbf{e}_1   + 1.0 \mathbf{e}_2 \otimes \mathbf{e}_2   + 1.0 \mathbf{e}_3 \otimes \mathbf{e}_3  $, \\
\hline
\end{tabular}
\caption{O3}
\end{subtable}
\caption{Invariant and exclusive 2nd order tensors for \SOthree\
(\tref{tab:crystal_groups}a) and \Othree\
(\tref{tab:crystal_groups}b) groups.}
\label{tab:2nd_order}
\end{table}

\paragraph{Fourth order structure tensors}

As can be seen in \tref{tab:crystal_groups}, fourth order tensors provide at least one characteristic tensor for each crystal subgroup of \Othree\ as defined by \tref{tab:crystal_groups}b.
Furthermore, the expected embedding is preserved by the bases (prior to eliminating the non-exclusive basis elements).
The components for all the bases (except monoclinic which is omitted for brevity) are given in \aref{app:4th_order}.

We used the in-span test, Eq. \eqref{eq:inspan}, to confirm that the bases found by the algorithm in \sref{sec:method} contain known fourth order structure tensors.
The isotropic fourth order tensors are well-known:
$\Ibb$, $\Jbb$, and $\Kbb$, see Eq. \eqref{eq:IJK}. 
which have the same span as the 3 bases we find and are given in \tref{tab:isotropic}.
The cubic group is characterized by $\Obb_h$ \eqref{eq:O}
which is identical to one of the bases we find and are enumerated in \tref{tab:cubic}.
Likewise, the tetragonal (dihedral 4) group has a structural tensor $\Dbb_{4h}$ \eqref{eq:D},
which is in the span of the basis we find and, in fact, has non-zero contributions from the 8 basis elements in \tref{tab:tetragonal}.

Lastly, \tref{tab:single_4th_order} enumerates the single simplest structural tensors for each group using \Aref{alg:simplify}.

\begin{table}[h]
\centering
\begin{subtable}[b]{0.49\textwidth}
\centering
\begin{tabular}{|c|cccc|l|}
\hline
Group & \multicolumn{4}{c|}{indices} &  value \\
\hline
monoclinic & 1 & 1 & 2 & 1 &   1.0000 \\
\hline
orthotropic & 2 & 2 & 3 & 3 &   1.0000 \\
\hline
trigonal & 3 & 1 & 1 & 3 &   1.0000 \\
& 3 & 2 & 2 & 3 &   1.0000 \\
\hline
tetragonal & 1 & 1 & 2 & 2 &   1.0000 \\
& 2 & 2 & 1 & 1 &   1.0000 \\
\hline
cubic & 1 & 1 & 1 & 1 &   1.0000 \\
& 2 & 2 & 2 & 2 &   1.0000 \\
& 3 & 3 & 3 & 3 &   1.0000 \\
\hline
transverse & 1 & 3 & 1 & 3 &   1.0000 \\
& 2 & 3 & 2 & 3 &   1.0000 \\
\hline
isotropic & 1 & 1 & 1 & 1 &   1.0000 \\
& 1 & 1 & 2 & 2 &   1.0000 \\
& 1 & 1 & 3 & 3 &   1.0000 \\
& 2 & 2 & 1 & 1 &   1.0000 \\
& 2 & 2 & 2 & 2 &   1.0000 \\
& 2 & 2 & 3 & 3 &   1.0000 \\
& 3 & 3 & 1 & 1 &   1.0000 \\
& 3 & 3 & 2 & 2 &   1.0000 \\
& 3 & 3 & 3 & 3 &   1.0000 \\
\hline
\end{tabular}
\caption{SO3}
\end{subtable}
\begin{subtable}[b]{0.49\textwidth}
\centering
\begin{tabular}{|c|cccc|l|}
\hline
Group & \multicolumn{4}{c|}{indices} &  value \\
\hline
monoclinic & 1 & 3 & 3 & 1 &   1.0000 \\
\hline
orthotropic & 2 & 2 & 3 & 3 &   1.0000 \\
\hline
trigonal & 3 & 1 & 3 & 1 &   1.0000 \\
& 3 & 2 & 3 & 2 &   1.0000 \\
\hline
tetragonal & 1 & 3 & 1 & 3 &   1.0000 \\
& 2 & 3 & 2 & 3 &   1.0000 \\
\hline
cubic & 1 & 1 & 1 & 1 &   1.0000 \\
& 2 & 2 & 2 & 2 &   1.0000 \\
& 3 & 3 & 3 & 3 &   1.0000 \\
\hline
transverse & 1 & 2 & 1 & 2 &  -1.0000 \\
& 1 & 2 & 2 & 1 &   1.0000 \\
& 2 & 1 & 1 & 2 &   1.0000 \\
& 2 & 1 & 2 & 1 &  -1.0000 \\
\hline
isotropic & 1 & 1 & 1 & 1 &   1.0000 \\
& 1 & 1 & 2 & 2 &   1.0000 \\
& 1 & 1 & 3 & 3 &   1.0000 \\
& 2 & 2 & 1 & 1 &   1.0000 \\
& 2 & 2 & 2 & 2 &   1.0000 \\
& 2 & 2 & 3 & 3 &   1.0000 \\
& 3 & 3 & 1 & 1 &   1.0000 \\
& 3 & 3 & 2 & 2 &   1.0000 \\
& 3 & 3 & 3 & 3 &   1.0000 \\
\hline
\end{tabular}
\caption{O3}
\end{subtable}
\caption{Components of single simple fourthorder tensors.
The components for the triclinic group are omitted for brevity.}
\label{tab:single_4th_order}
\end{table}

\subsection{Learning elastic anisotropy class and orientation}

The results of \sref{sec:crystal_groups} can be used in learning anisotropy in constitutive response from stress-strain data, as in the sparse regression approaches in\cref{fuhg2022learning} and \cref{jadoon2025inverse}.
Kalina et al. \cite{kalina2025neural} use a similar approach but in a step-by-step learning process that incrementally increases model complexity, allowing anisotropy and its orientation to come from data through a trial-and-error refinement loop by training different neural networks for different symmetry orders.
In contrast, the proposed framework, for the first time, provides a full parameterization of all symmetry classes {\it a priori} using a fixed basis of structural tensors and sparse regression to select the appropriate anisotropy from the entire symmetry space.

Here, we use the single, simple fourth order structure tensors for all groups enumerated in
\tref{tab:single_4th_order} to obtain a set of possible structure tensors
\begin{equation}
\{ \structuraltensor_a \} = \{ \structuraltensor_\text{triclinic}, \structuraltensor_\text{orthorhombic}, \ldots, \structuraltensor_\text{isotropic} \}
\end{equation}
since each group has a single fourth order structure tensor.
This allows us to use a common form for the joint invariants of the strain $\strain$ and any structural tensor $\Abb$:
\begin{equation}
\stress = \partialb_\strain \Psi(\strain, \structuraltensor)
= \partialb_\strain \Psi(\Ic_{\strain,\structuraltensor})
= \sum_b \partialb_{I_b} \Psi \, \partialb_\strain I_b
= \sum_b c_b (I_b) \, \Bb_b
\quad
\text{where} \ I_b \in \Ic_{\strain,\structuraltensor}
\end{equation}
For a single fourth order structural tensor, Kambouchev \etal \cite{kambouchev2007polyconvex} provided the invariants for a polyconvex hyperelastic model
\begin{equation} \label{eq:kam_invar}
\Ic_{\strain,\structuraltensor}
= \{ \underbrace{\tr \strain, \tr \strain^*, \det \strain}_{\invariants_\text{iso}},
\underbrace{
\structuraltensor:\strain\otimes\strain,
\structuraltensor:\strain \otimes \strain^2,
\structuraltensor: \strain^2 \otimes \strain^2}_{\invariants_\structuraltensor}
\}
\end{equation}
where the anisotropic invariants $\invariants_\structuraltensor$ are traces of products of $\structuraltensor$ with powers of $\strain$.
(Note Kambouchev \etal \cite{kambouchev2007polyconvex} was restricted to the cubic structure tensor $\Obb_h$.)
A corresponding tensor basis can be obtained from the derivatives of the invariants:
\begin{equation} \label{eq:kam_basis}
\basis = \{ \partial_\strain \invariants_i \} = \{ \underbrace{ \Bb_1\tighteq \Ib, \Bb_2\tighteq \strain, \Bb_3\tighteq \strain^* }_{\basis_\text{iso}} \} \cup \{ \underbrace{ \Bb_4, \Bb_5, \Bb_6 }_{\basis_\Abb} \}
\end{equation}
where
\begin{eqnarray}
\Bb_4 &=& ( \structuraltensor_{IJKL} \strain_{KL} + \structuraltensor_{KLIJ} \strain_{KL} ) \eb_I \otimes \eb_J, \\
\Bb_5 &=& \biggl(
(\structuraltensor_{IJKL} + \structuraltensor_{KLIJ}) \strain^2_{KL}, \\
&&+
(\structuraltensor_{KLIQ} + \structuraltensor_{IQKL}) \strain_{KL} \strain_{JQ} +
(\structuraltensor_{KLQJ} + \structuraltensor_{QJKL}) \strain_{KL} \strain_{QI}
\biggr) \eb_I \otimes \eb_J  \nonumber \\
\Bb_6 &=& 2 \biggl(
(\structuraltensor_{KLIQ} + \structuraltensor_{IQKL}) \strain^2_{KL} \strain_{JQ} +
(\structuraltensor_{KLQJ} + \structuraltensor_{QJKL}) \strain^2_{KL} \strain_{QI}
\biggr) \eb_I \otimes \eb_J  .
\end{eqnarray}

As in \cref{fuhg2022learning} we also discover the orientation $\Qb \in$ \SOthree\ of the structure tensor relative to the canonically oriented structure tensors $\{\structuraltensor_a\}$ as:
\begin{equation}
\structuraltensor'
= \rotation \boxtimes \structuraltensor
= \rotation \boxtimes \sum_a c_a \structuraltensor_a
\end{equation}
The rotation $\rotation$, has a convenient Euler-Rodrigues representation \eqref{eq:rodriguez} in terms of a rotation angle $\theta\in [0,2\pi)$ and a (unit) axis vector $\pb$.
In model training,  we infer a 3 vector and take its magnitude to be $\theta$ and its direction to be $\pb$.

We form the trial potential energy from $\Ic_\text{iso}$ and the \emph{independent} contributions of each structural tensor $\Ic_{\structuraltensor_a}, a \in \grp$
\begin{equation}
\Psi = \Psi(\Ic_\text{iso}, \alpha_1 \Ic_1, \alpha_2 \Ic_2, \ldots, \alpha_n \Ic_n)
\end{equation}
which is extensible to joint invariants, albeit with a significant expansion of the inputs.
We tie the invariants of each group together with a common weight $\alpha_a$.
Alternatively, we could have used an ansatz formed from individual energies:
\begin{equation}
\Psi = \sum_a \alpha_a \Psi(\Ic_\text{iso}, \Ic_a)  \quad | \quad \alpha_a > 0
\end{equation}
or
a formulation that penalizes each basis contribution
\begin{equation}
\stress = \sum_a \alpha_a c_a \Bb_a \ .
\end{equation}
For this demonstration, we want to find the set of structural tensors $\Ac = \{ \alpha_a \Ab_a \}$ such that the active ones ($\alpha_a > 0$) represent the data symmetries.
Here we use the fourth order single tensors listed in \tref{tab:single_4th_order}.
Note that the structural tensors in \tref{tab:single_4th_order} are not orthogonal, so combinations of them are possible solutions.
(A Graham-Schmidt procedure could be used to orthogonalize the set but the association with specific groups would be lost.)
We augment the usual mean squared error loss with a L1 norm of the coefficients $\alphab = (\alpha_a, a=1,n)$ to enable sparse regression
\begin{equation}
\Lc = \| \Sb - \partialb_\Eb \Psi(\Eb, \alpha_a \structuraltensor_a)\|_2
+ \lambda \| \alphab \|_1
\end{equation}

We generated data for a transversely isotropic material with symmetry axis $\pb = (\eb_1 + \eb_2)/\sqrt{2}$ using a Bonet-Burton hyperelastic model \cite{bonet1998simple}, as in \cref{fuhg2022learning}.
A dataset of 1000 samples, generated by a random sample of deformation space up to 0.4 strain, was divided into 800 training and 200 testing samples.

\begin{figure}
\centering
\includegraphics[width=0.5\linewidth]{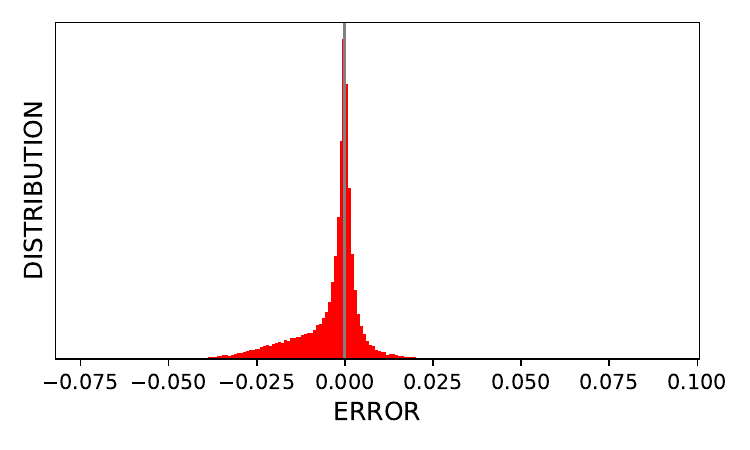}
\caption{Anisotropic elasticity: model-data component difference error distribution.}
\label{fig:aniso_elast-error}
\end{figure}

\begin{figure}
\centering
\includegraphics[width=0.5\linewidth]{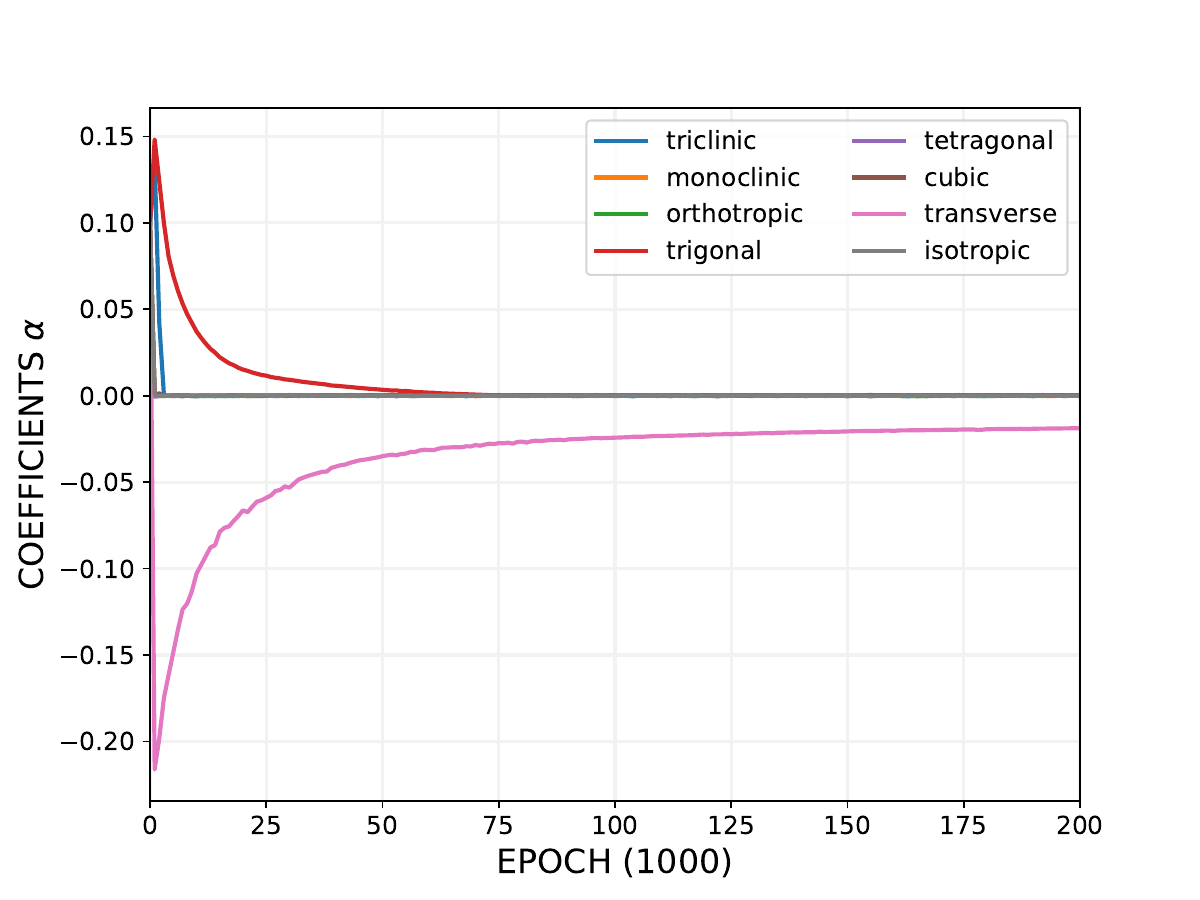}
\caption{Anisotropic elasticity: structure tensor (sparsity) weights $\alpha_a$ where
1:triclinic, 2: monoclinic, 3:orthotropic, 4:trigonal, 5: tetragonal, 6:cubic, 7: transverse, 8: isotropic.}
\label{fig:ansio_elast-weights}
\end{figure}

\begin{figure}
\centering
\begin{subfigure}[b]{0.45\linewidth}
\includegraphics[width=0.95\linewidth]{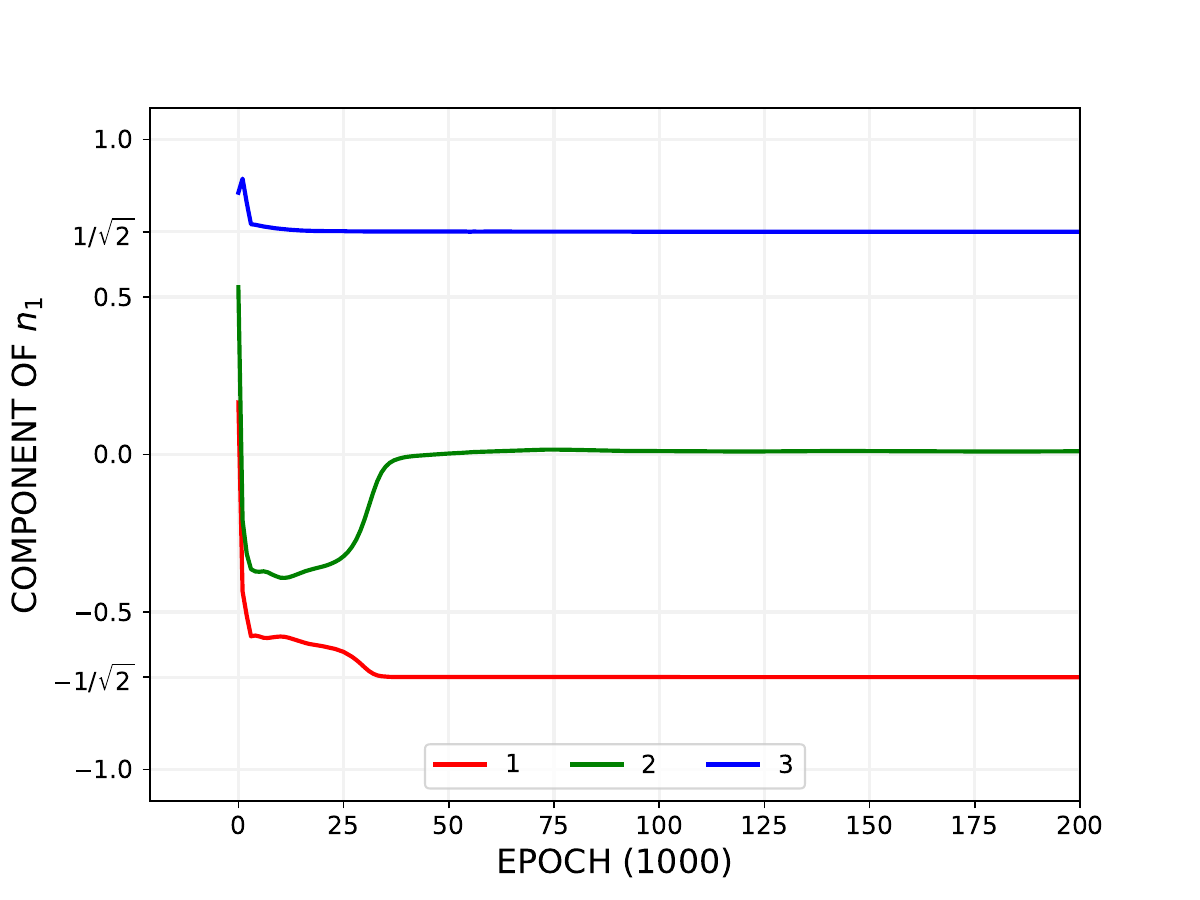}
\caption{components of $\nb_1$}
\end{subfigure}
\begin{subfigure}[b]{0.45\linewidth}
\includegraphics[width=0.95\linewidth]{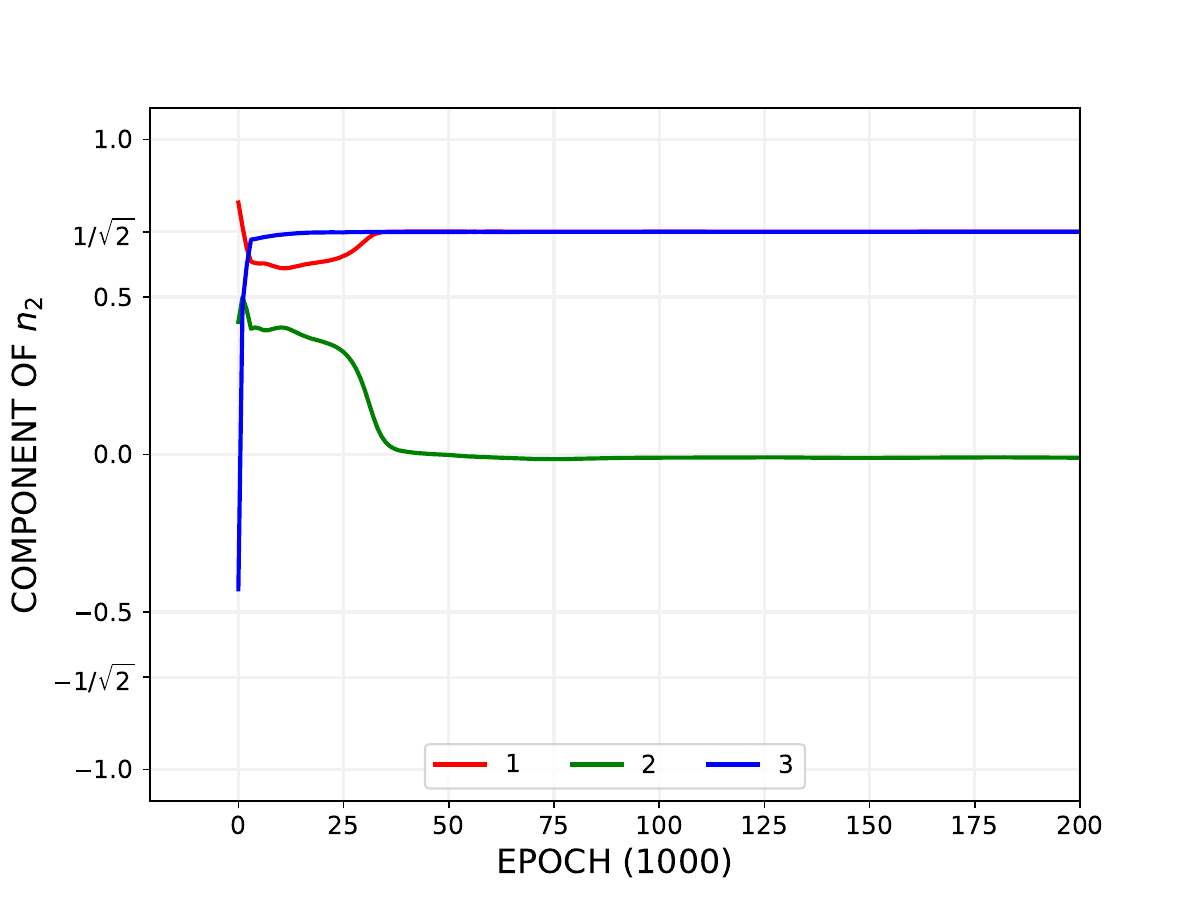}
\caption{components of $\nb_2$}
\end{subfigure}
\caption{Anisotropic elasticity: orientation
triad $\nb_1,\nb_2,\nb_3=\nb_1\times\nb_2$.}
\label{fig:aniso_elast-orientation}
\end{figure}

For the potential, we employed an input convex neural network (ICNN) \cite{amos2017input} with inputs: the 6 invariants of each group, 3 hidden layers with 30 neurons each, and a single output: the value of the potential.
The model achieves a root mean squared error (RMSE) of  0.00983 (0.0101 relative to normalized data) and a correlation of  1.0000 on held-out data.
\fref{fig:aniso_elast-error} indicates that the accuracy of predictions of the model on held-out data is excellent with the mode in the error distribution at zero, albeit with a slightly different decay to over versus under-predict stress.
\fref{fig:ansio_elast-weights} shows the evolution of sparsity coefficients $\alpha_a$ during training.
After some transients in the gradient descent procedure (ADAM \cite{kingma2014adam}), the structure tensor $\structuraltensor_\text{transverse}$, which corresponds to transverse isotropic symmetries, dominates,  so that it is clear the method discovers the correct symmetry from the stress-strain data.
\fref{fig:aniso_elast-orientation} shows that the correct orientation is also recovered.

\subsection{Learning strain-dependent elastic anisotropy}

Unlike in the previous example, here we use the basis from the lowest symmetry group we expect, which can represent all the higher symmetries in the chain of embedded groups in poset \fref{fig:SO3_poset}.
(In this context a group has higher symmetry than another if its generators subsume the generators of the lower symmetry group.)
Again, we seek to learn the members of the set that are needed to represent the data response.

\begin{figure}
\centering
\begin{subfigure}[b]{0.49\textwidth}
\centering
\includegraphics[width=0.7\linewidth]{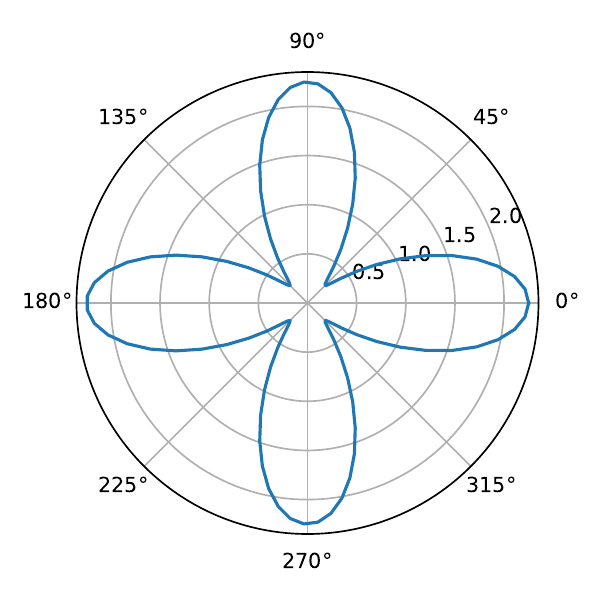}
\caption{Distribution}
\end{subfigure}
\begin{subfigure}[b]{0.49\textwidth}
\centering
\includegraphics[width=0.7\linewidth]{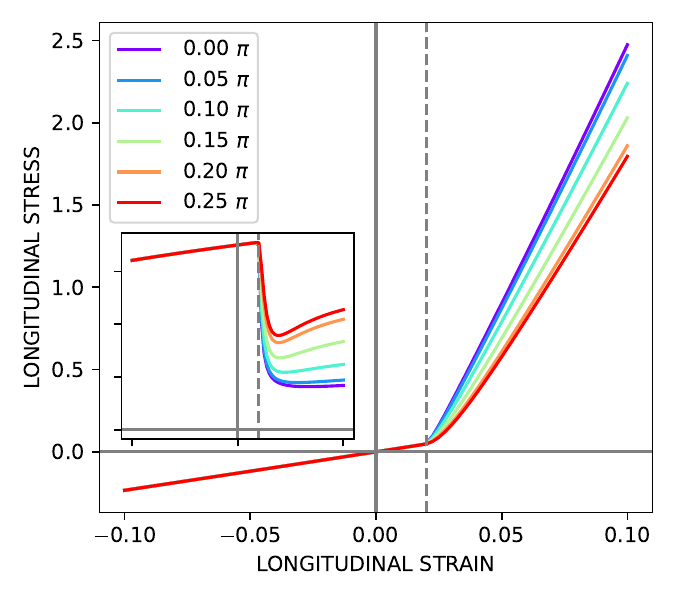}
\caption{Stress}
\end{subfigure}
\caption{Fiber (a) distribution and (b) response. The inset of panel (b) shows the ratio of perpendicular to longitudinal (parallel) stress.}
\label{fig:fiber_stress}
\end{figure}

For this demonstration, we employ a data-generating model that displays
strain-dependent anisotropy evocative of a composite with a soft matrix embedded with strong but loose fibers.
Near the reference configuration, the material behaves isotropically, but anisotropy becomes more pronounced as the fibers are stretched.
Specifically, the strain energy is given in a model form akin to \crefs{gasser2006hyperelastic,pandolfi2012fiber}:
\begin{equation}
\Psi = \Psi_\text{bulk}(\Cb) + \int_0^{2\pi} \psi(\lambda(\theta)) \,
\mathrm{d}\rho(\theta)  \ ,
\end{equation}
where the stretch is $\lambda = \sqrt{\Nb(\theta) \cdot \Cb \Nb(\theta)}$ and a neo-Hookean model gives the bulk response.
The corresponding stress is
\begin{equation}
\stress = \partialb_{\strain} \Psi
= \stress_\text{bulk} +
\int_0^{2\pi} \psi'(\lambda(\theta))
\Nb(\theta) \otimes
\Nb(\theta) \,
\mathrm{d}\rho(\theta)
\end{equation}
where $\Nb(\theta)$ is a vector tangent to in-plane fibers with density $\rho(\theta)$ and out-of-plane direction $\eb_3$.
The distribution of fibers has a 4-fold symmetry:
\begin{equation}
\rho(\theta) =  \tfrac{1}{4}+ (\cos(4 \theta) + 1) \ ,
\end{equation}
see \fref{fig:fiber_stress}a.
To emulate loose, stiff fibers with no compressive strength in a soft matrix we use
\begin{equation}
\psi = \begin{cases}
E (\lambda-\lambda_c)^2 & \text{if} \ \  \lambda > \lambda_c \\
0 & \text{else}
\end{cases}
\end{equation}
where $E$ is the fiber modulus and $\lambda_c$ is the critical stretch.

We presume that the rough symmetry of the fiber distribution in \fref{fig:fiber_stress}a is from microstructure observations, so that the stress-strain experiments do not sample beyond the presumed tetragonal symmetries.
Specifically, we collect uniaxial stretch data on 16 equally spaced orientations $\theta = [0,\pi/4)$ (relative to a reference orientation for the material).
\fref{fig:fiber_stress} shows the stress response for these loadings.
Clearly, the anisotropy activates at a strain of 0.02 and increases with increasing tension.
We validate the model on held-out orientations that should be represented by symmetry operations.

For this model, we make the coefficients $\alpha_a$ functions of the invariants
\begin{equation}
\alpha_a = \NN_\alpha(\Ic_{\structuraltensor_a})
\end{equation}
so that the stress model is
\begin{equation}
\Sb = \partialb_{\strain} \NN_\Psi(\strain, \NN_\alpha(\strain) \structuraltensor_a)
\end{equation}
where $\NN_\Psi$ is an input convex NN \cite{amos2017input,klein2023parametrized,fuhg2024polyconvex} and each $\NN_\alpha(\strain)$ has a sigmoid output to activate/deactivate anisotropy based on deformation.
Given the symmetry of the observed fiber distribution, we employed the tetragonal basis enumerated in \tref{tab:tetragonal}.

\begin{figure}
\centering
\begin{subfigure}{0.85\linewidth}
\includegraphics[width=0.99\linewidth]{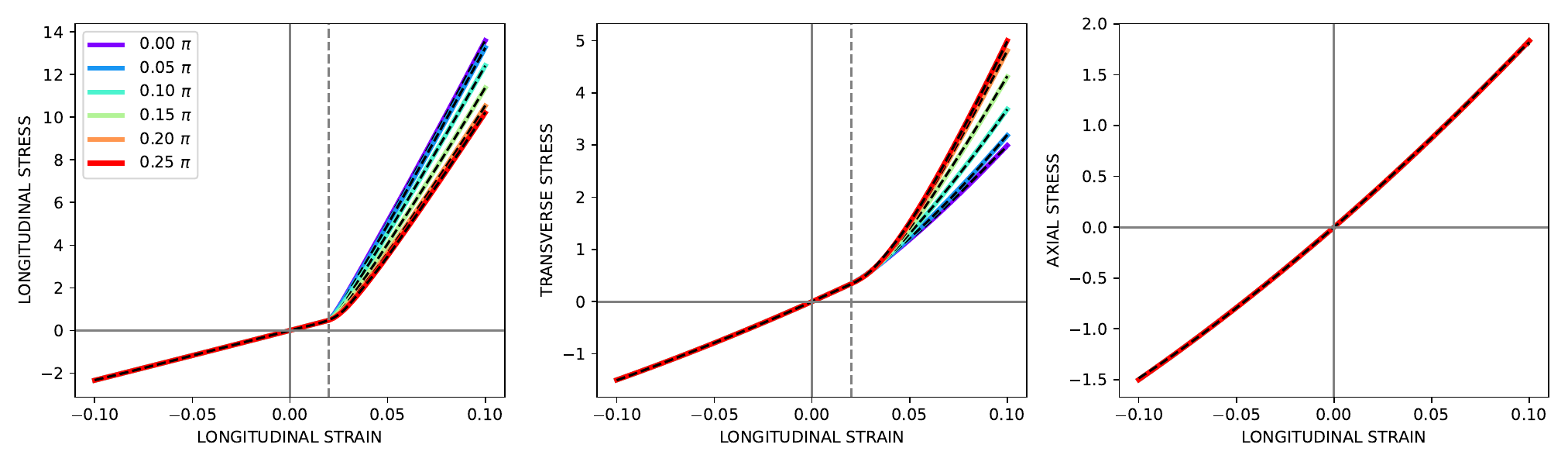}
\caption{training}
\end{subfigure}
\begin{subfigure}{0.85\linewidth}
\includegraphics[width=0.99\linewidth]{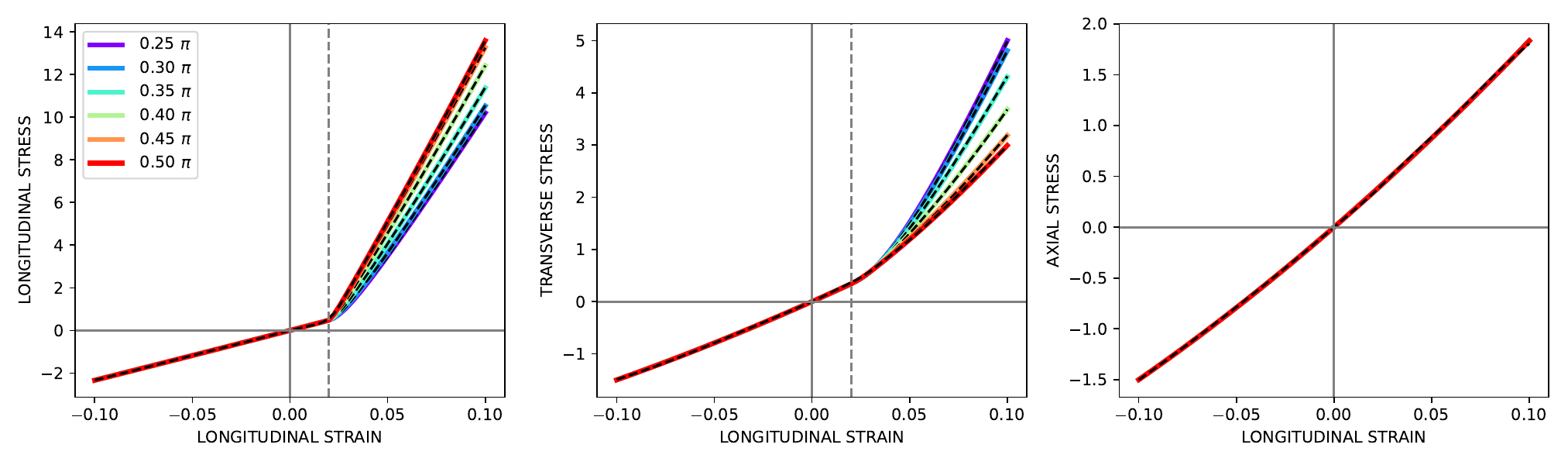}
\caption{test}
\end{subfigure}
\caption{Fiber stress comparison for unixaial tests at different orientations, data (color lines), model (dashed black lines).
Left to right: stress parallel to in-plane uniaxial loading, stress perpendicular loading, and out-of-plane stress.
(a) training data, (b) held-out testing data demonstrating the symmetries are preserved.
}
\label{fig:fiber_stress_comparison}
\end{figure}

\fref{fig:fiber_stress_comparison}a shows that the model trains well and represents the strain-dependent anisotropy well.
The essentially identical results in \fref{fig:fiber_stress_comparison}b indicate that the symmetries and accuracy are preserved on held-out data.

Note that a separate calibration using the single $\Dbb_{4h}$  structure tensor did not produce a low discrepancy model, which we attribute to the potential energy being a sum of different planar and bulk contributions.
The same formulation trained perfectly to a corresponding dataset with the fiber response generated directly from $\Dbb_{4h}$.
This finding emphasizes the value of calibrating with a full basis of structural tensors.

\section{Conclusion} \label{sec:conclusion}

We have proposed a simple method to generate invariant tensors of a given degree given the (orthogonal) generators of the symmetry group, any permutational/transpositional symmetries, and any other constraints that are linear in the tensor itself.
Both the rigorous mathematical foundation and simple to implement algorithms were provided.
Furthermore, given the methods' reliance on the SVD, they are differentiable algorithms, which can be useful in model calibration and optimal experimental design.
We have shown how this constructive method can be used in ubiquitous modeling tasks in mechanics such as forming a basis for moduli and constructing structural tensors.
In fact, the method produces a basis for invariant tensors for any given group, and a simple algorithm can be used to find a single invariant tensor with the fewest non-zero components.
We demonstrated how calibration to unknown symmetries can be performed using stress-strain data and either: (a) single characteristic tensors from each possible symmetry group, or (b) a basis of invariant tensors from the lowest expected symmetry group.
In the sparse regression, neural network framework we presented, this enables the discovery of the appropriate symmetry class and representation of the response function to a high degree of accuracy.
The utility of the method in function representation does rely on other tools such as isotropic function representations and formulae for joint invariants.
Furthermore, we currently need to rely on subgroup ordering to test whether basis elements are exclusive of other (orthogonal) subgroups.

In future work, we want to apply the method to plastically evolving anisotropy due to work-induced texture changes \cite{arruda1993evolution,ahzi1994plasticity,hiwatashi1997modelling,schiferl1997evolution,li2003finite,kowalczyk2004model,harrysson2007description,kweon2017comparison,negahban1992evolution,rajagopal1992constitutive}.
Also the method should prove useful in discovering the symmetries and best representation of the response of highly textured crystal plasticity samples and other representative volume element of materials with pronounced microstructure.
While we provide some empirical evidence that the sparsity used in Alg.~\ref{alg:alg2} is sufficient to find the image of the Reynolds operator, we will try to obtain probabilistic bounds.
Moreover, we suspect this randomized algorithm provides a route for accelerating algorithms for obtaining the integrity basis for invariant functions because such algorithms often require one to obtain the invariant subspaces of symmetric powers \cite{derksen2015computational}.
Future work will investigate this route towards accelerating these algorithms, which is indispensable for higher-order tensor applications.

\bibliographystyle{unsrt}


\appendix

\setcounter{equation}{0}
\setcounter{table}{0}

\raggedbottom
\afterpage{\clearpage}

\clearpage
\renewcommand{\theequation}{\thesection-\arabic{equation}}
\renewcommand{\thetable}{\thesection-\arabic{table}}

\setcounter{equation}{0}
\setcounter{table}{0}
`\section{Components of elastic modulus tensors for common symmetry groups} \label{app:elastic_modulus}
\normalsize
The following tables enumerate the bases for the elastic modulus tensors for isotropic \tref{tab:isotropic elastic modulus}, cubic \tref{tab:cubic elastic modulus}, tetragonal \tref{tab:tetragonal elastic modulus}, and orthotropic \tref{tab:orthotropic elastic modulus} materials.

\begin{table}
\centering
\begin{subtable}[c]{0.3\textwidth}
\begin{tabular}{|cccc|c|}
\hline
1 & 1 & 1 & 1 &  -0.0325 \\
1 & 1 & 2 & 2 &   1.0000 \\
1 & 1 & 3 & 3 &   1.0000 \\
1 & 2 & 1 & 2 &  -0.5163 \\
1 & 2 & 2 & 1 &  -0.5163 \\
1 & 3 & 1 & 3 &  -0.5163 \\
1 & 3 & 3 & 1 &  -0.5163 \\
2 & 1 & 1 & 2 &  -0.5163 \\
2 & 1 & 2 & 1 &  -0.5163 \\
2 & 2 & 1 & 1 &   1.0000 \\
2 & 2 & 2 & 2 &  -0.0325 \\
2 & 2 & 3 & 3 &   1.0000 \\
2 & 3 & 2 & 3 &  -0.5163 \\
2 & 3 & 3 & 2 &  -0.5163 \\
3 & 1 & 1 & 3 &  -0.5163 \\
3 & 1 & 3 & 1 &  -0.5163 \\
3 & 2 & 2 & 3 &  -0.5163 \\
3 & 2 & 3 & 2 &  -0.5163 \\
3 & 3 & 1 & 1 &   1.0000 \\
3 & 3 & 2 & 2 &   1.0000 \\
3 & 3 & 3 & 3 &  -0.0325 \\
\hline
\end{tabular}
\caption{base 1}
\end{subtable}
\begin{subtable}[c]{0.3\textwidth}
\begin{tabular}{|cccc|c|}
\hline
1 & 1 & 1 & 1 &   1.0000 \\
1 & 1 & 2 & 2 &   0.3512 \\
1 & 1 & 3 & 3 &   0.3512 \\
1 & 2 & 1 & 2 &   0.3244 \\
1 & 2 & 2 & 1 &   0.3244 \\
1 & 3 & 1 & 3 &   0.3244 \\
1 & 3 & 3 & 1 &   0.3244 \\
2 & 1 & 1 & 2 &   0.3244 \\
2 & 1 & 2 & 1 &   0.3244 \\
2 & 2 & 1 & 1 &   0.3512 \\
2 & 2 & 2 & 2 &   1.0000 \\
2 & 2 & 3 & 3 &   0.3512 \\
2 & 3 & 2 & 3 &   0.3244 \\
2 & 3 & 3 & 2 &   0.3244 \\
3 & 1 & 1 & 3 &   0.3244 \\
3 & 1 & 3 & 1 &   0.3244 \\
3 & 2 & 2 & 3 &   0.3244 \\
3 & 2 & 3 & 2 &   0.3244 \\
3 & 3 & 1 & 1 &   0.3512 \\
3 & 3 & 2 & 2 &   0.3512 \\
3 & 3 & 3 & 3 &   1.0000 \\
\hline
\end{tabular}
\caption{base 2}
\end{subtable}
\caption{ isotropic elastic modulus: 2 bases }
\label{tab:isotropic elastic modulus}
\end{table}
\afterpage{\clearpage}

\begin{table}
\centering
\begin{subtable}[c]{0.3\textwidth}
\begin{tabular}{|cccc|c|}
\hline
1 & 1 & 1 & 1 &   1.0000 \\
2 & 2 & 2 & 2 &   1.0000 \\
3 & 3 & 3 & 3 &   1.0000 \\
\hline
\end{tabular}
\caption{base 1}
\end{subtable}
\begin{subtable}[c]{0.3\textwidth}
\begin{tabular}{|cccc|c|}
\hline
1 & 1 & 2 & 2 &   0.6992 \\
1 & 1 & 3 & 3 &   0.6992 \\
1 & 2 & 1 & 2 &   1.0000 \\
1 & 2 & 2 & 1 &   1.0000 \\
1 & 3 & 1 & 3 &   1.0000 \\
1 & 3 & 3 & 1 &   1.0000 \\
2 & 1 & 1 & 2 &   1.0000 \\
2 & 1 & 2 & 1 &   1.0000 \\
2 & 2 & 1 & 1 &   0.6992 \\
2 & 2 & 3 & 3 &   0.6992 \\
2 & 3 & 2 & 3 &   1.0000 \\
2 & 3 & 3 & 2 &   1.0000 \\
3 & 1 & 1 & 3 &   1.0000 \\
3 & 1 & 3 & 1 &   1.0000 \\
3 & 2 & 2 & 3 &   1.0000 \\
3 & 2 & 3 & 2 &   1.0000 \\
3 & 3 & 1 & 1 &   0.6992 \\
3 & 3 & 2 & 2 &   0.6992 \\
\hline
\end{tabular}
\caption{base 2}
\end{subtable}
\begin{subtable}[c]{0.3\textwidth}
\begin{tabular}{|cccc|c|}
\hline
1 & 1 & 2 & 2 &   1.0000 \\
1 & 1 & 3 & 3 &   1.0000 \\
1 & 2 & 1 & 2 &  -0.3496 \\
1 & 2 & 2 & 1 &  -0.3496 \\
1 & 3 & 1 & 3 &  -0.3496 \\
1 & 3 & 3 & 1 &  -0.3496 \\
2 & 1 & 1 & 2 &  -0.3496 \\
2 & 1 & 2 & 1 &  -0.3496 \\
2 & 2 & 1 & 1 &   1.0000 \\
2 & 2 & 3 & 3 &   1.0000 \\
2 & 3 & 2 & 3 &  -0.3496 \\
2 & 3 & 3 & 2 &  -0.3496 \\
3 & 1 & 1 & 3 &  -0.3496 \\
3 & 1 & 3 & 1 &  -0.3496 \\
3 & 2 & 2 & 3 &  -0.3496 \\
3 & 2 & 3 & 2 &  -0.3496 \\
3 & 3 & 1 & 1 &   1.0000 \\
3 & 3 & 2 & 2 &   1.0000 \\
\hline
\end{tabular}
\caption{base 3}
\end{subtable}
\caption{ cubic elastic modulus: 3 bases }
\label{tab:cubic elastic modulus}
\end{table}
\afterpage{\clearpage}

\begin{table}
\centering
\begin{subtable}[c]{0.3\textwidth}
\begin{tabular}{|cccc|c|}
\hline
3 & 3 & 3 & 3 &   1.0000 \\
\hline
\end{tabular}
\caption{base 1}
\end{subtable}
\begin{subtable}[c]{0.3\textwidth}
\begin{tabular}{|cccc|c|}
\hline
1 & 1 & 1 & 1 &   1.0000 \\
2 & 2 & 2 & 2 &   1.0000 \\
\hline
\end{tabular}
\caption{base 2}
\end{subtable}
\begin{subtable}[c]{0.3\textwidth}
\begin{tabular}{|cccc|c|}
\hline
1 & 1 & 2 & 2 &   1.0000 \\
1 & 1 & 3 & 3 &  -0.4538 \\
1 & 2 & 1 & 2 &   0.3225 \\
1 & 2 & 2 & 1 &   0.3225 \\
1 & 3 & 1 & 3 &  -0.0676 \\
1 & 3 & 3 & 1 &  -0.0676 \\
2 & 1 & 1 & 2 &   0.3225 \\
2 & 1 & 2 & 1 &   0.3225 \\
2 & 2 & 1 & 1 &   1.0000 \\
2 & 2 & 3 & 3 &  -0.4538 \\
2 & 3 & 2 & 3 &  -0.0676 \\
2 & 3 & 3 & 2 &  -0.0676 \\
3 & 1 & 1 & 3 &  -0.0676 \\
3 & 1 & 3 & 1 &  -0.0676 \\
3 & 2 & 2 & 3 &  -0.0676 \\
3 & 2 & 3 & 2 &  -0.0676 \\
3 & 3 & 1 & 1 &  -0.4538 \\
3 & 3 & 2 & 2 &  -0.4538 \\
\hline
\end{tabular}
\caption{base 3}
\end{subtable}
\begin{subtable}[c]{0.3\textwidth}
\begin{tabular}{|cccc|c|}
\hline
1 & 1 & 2 & 2 &   0.7981 \\
1 & 1 & 3 & 3 &   1.0000 \\
1 & 2 & 1 & 2 &  -0.0368 \\
1 & 2 & 2 & 1 &  -0.0368 \\
1 & 3 & 1 & 3 &  -0.4923 \\
1 & 3 & 3 & 1 &  -0.4923 \\
2 & 1 & 1 & 2 &  -0.0368 \\
2 & 1 & 2 & 1 &  -0.0368 \\
2 & 2 & 1 & 1 &   0.7981 \\
2 & 2 & 3 & 3 &   1.0000 \\
2 & 3 & 2 & 3 &  -0.4923 \\
2 & 3 & 3 & 2 &  -0.4923 \\
3 & 1 & 1 & 3 &  -0.4923 \\
3 & 1 & 3 & 1 &  -0.4923 \\
3 & 2 & 2 & 3 &  -0.4923 \\
3 & 2 & 3 & 2 &  -0.4923 \\
3 & 3 & 1 & 1 &   1.0000 \\
3 & 3 & 2 & 2 &   1.0000 \\
\hline
\end{tabular}
\caption{base 4}
\end{subtable}
\begin{subtable}[c]{0.3\textwidth}
\begin{tabular}{|cccc|c|}
\hline
1 & 1 & 2 & 2 &  -0.6806 \\
1 & 1 & 3 & 3 &   0.0415 \\
1 & 2 & 1 & 2 &   1.0000 \\
1 & 2 & 2 & 1 &   1.0000 \\
1 & 3 & 1 & 3 &  -0.2711 \\
1 & 3 & 3 & 1 &  -0.2711 \\
2 & 1 & 1 & 2 &   1.0000 \\
2 & 1 & 2 & 1 &   1.0000 \\
2 & 2 & 1 & 1 &  -0.6806 \\
2 & 2 & 3 & 3 &   0.0415 \\
2 & 3 & 2 & 3 &  -0.2711 \\
2 & 3 & 3 & 2 &  -0.2711 \\
3 & 1 & 1 & 3 &  -0.2711 \\
3 & 1 & 3 & 1 &  -0.2711 \\
3 & 2 & 2 & 3 &  -0.2711 \\
3 & 2 & 3 & 2 &  -0.2711 \\
3 & 3 & 1 & 1 &   0.0415 \\
3 & 3 & 2 & 2 &   0.0415 \\
\hline
\end{tabular}
\caption{base 5}
\end{subtable}
\begin{subtable}[c]{0.3\textwidth}
\begin{tabular}{|cccc|c|}
\hline
1 & 1 & 2 & 2 &   0.5429 \\
1 & 1 & 3 & 3 &   0.7934 \\
1 & 2 & 1 & 2 &   0.6941 \\
1 & 2 & 2 & 1 &   0.6941 \\
1 & 3 & 1 & 3 &   1.0000 \\
1 & 3 & 3 & 1 &   1.0000 \\
2 & 1 & 1 & 2 &   0.6941 \\
2 & 1 & 2 & 1 &   0.6941 \\
2 & 2 & 1 & 1 &   0.5429 \\
2 & 2 & 3 & 3 &   0.7934 \\
2 & 3 & 2 & 3 &   1.0000 \\
2 & 3 & 3 & 2 &   1.0000 \\
3 & 1 & 1 & 3 &   1.0000 \\
3 & 1 & 3 & 1 &   1.0000 \\
3 & 2 & 2 & 3 &   1.0000 \\
3 & 2 & 3 & 2 &   1.0000 \\
3 & 3 & 1 & 1 &   0.7934 \\
3 & 3 & 2 & 2 &   0.7934 \\
\hline
\end{tabular}
\caption{base 6}
\end{subtable}
\caption{ tetragonal elastic modulus: 6 bases }
\label{tab:tetragonal elastic modulus}
\end{table}
\afterpage{\clearpage}

\begin{table}
\centering
\begin{subtable}[c]{0.3\textwidth}
\begin{tabular}{|cccc|c|}
\hline
3 & 3 & 3 & 3 &   1.0000 \\
\hline
\end{tabular}
\caption{base 1}
\end{subtable}
\begin{subtable}[c]{0.3\textwidth}
\begin{tabular}{|cccc|c|}
\hline
1 & 1 & 1 & 1 &   1.0000 \\
\hline
\end{tabular}
\caption{base 2}
\end{subtable}
\begin{subtable}[c]{0.3\textwidth}
\begin{tabular}{|cccc|c|}
\hline
2 & 2 & 2 & 2 &  -0.5262 \\
2 & 2 & 3 & 3 &   1.0000 \\
3 & 3 & 2 & 2 &   1.0000 \\
\hline
\end{tabular}
\caption{base 3}
\end{subtable}
\begin{subtable}[c]{0.3\textwidth}
\begin{tabular}{|cccc|c|}
\hline
2 & 2 & 2 & 2 &   1.0000 \\
2 & 2 & 3 & 3 &   0.2631 \\
3 & 3 & 2 & 2 &   0.2631 \\
\hline
\end{tabular}
\caption{base 4}
\end{subtable}
\begin{subtable}[c]{0.3\textwidth}
\begin{tabular}{|cccc|c|}
\hline
2 & 3 & 2 & 3 &   1.0000 \\
2 & 3 & 3 & 2 &   1.0000 \\
3 & 2 & 2 & 3 &   1.0000 \\
3 & 2 & 3 & 2 &   1.0000 \\
\hline
\end{tabular}
\caption{base 5}
\end{subtable}
\begin{subtable}[c]{0.3\textwidth}
\begin{tabular}{|cccc|c|}
\hline
1 & 1 & 2 & 2 &   1.0000 \\
1 & 1 & 3 & 3 &  -0.3478 \\
1 & 2 & 1 & 2 &  -0.4196 \\
1 & 2 & 2 & 1 &  -0.4196 \\
1 & 3 & 1 & 3 &   0.2294 \\
1 & 3 & 3 & 1 &   0.2294 \\
2 & 1 & 1 & 2 &  -0.4196 \\
2 & 1 & 2 & 1 &  -0.4196 \\
2 & 2 & 1 & 1 &   1.0000 \\
3 & 1 & 1 & 3 &   0.2294 \\
3 & 1 & 3 & 1 &   0.2294 \\
3 & 3 & 1 & 1 &  -0.3478 \\
\hline
\end{tabular}
\caption{base 6}
\end{subtable}
\begin{subtable}[c]{0.3\textwidth}
\begin{tabular}{|cccc|c|}
\hline
1 & 1 & 2 & 2 &  -0.3020 \\
1 & 1 & 3 & 3 &   1.0000 \\
1 & 2 & 1 & 2 &  -0.9188 \\
1 & 2 & 2 & 1 &  -0.9188 \\
1 & 3 & 1 & 3 &  -0.2646 \\
1 & 3 & 3 & 1 &  -0.2646 \\
2 & 1 & 1 & 2 &  -0.9188 \\
2 & 1 & 2 & 1 &  -0.9188 \\
2 & 2 & 1 & 1 &  -0.3020 \\
3 & 1 & 1 & 3 &  -0.2646 \\
3 & 1 & 3 & 1 &  -0.2646 \\
3 & 3 & 1 & 1 &   1.0000 \\
\hline
\end{tabular}
\caption{base 7}
\end{subtable}
\begin{subtable}[c]{0.3\textwidth}
\begin{tabular}{|cccc|c|}
\hline
1 & 1 & 2 & 2 &   0.9234 \\
1 & 1 & 3 & 3 &   1.0000 \\
1 & 2 & 1 & 2 &   0.4937 \\
1 & 2 & 2 & 1 &   0.4937 \\
1 & 3 & 1 & 3 &  -0.3516 \\
1 & 3 & 3 & 1 &  -0.3516 \\
2 & 1 & 1 & 2 &   0.4937 \\
2 & 1 & 2 & 1 &   0.4937 \\
2 & 2 & 1 & 1 &   0.9234 \\
3 & 1 & 1 & 3 &  -0.3516 \\
3 & 1 & 3 & 1 &  -0.3516 \\
3 & 3 & 1 & 1 &   1.0000 \\
\hline
\end{tabular}
\caption{base 8}
\end{subtable}
\begin{subtable}[c]{0.3\textwidth}
\begin{tabular}{|cccc|c|}
\hline
1 & 1 & 2 & 2 &  -0.1196 \\
1 & 1 & 3 & 3 &   0.7016 \\
1 & 2 & 1 & 2 &   0.1135 \\
1 & 2 & 2 & 1 &   0.1135 \\
1 & 3 & 1 & 3 &   1.0000 \\
1 & 3 & 3 & 1 &   1.0000 \\
2 & 1 & 1 & 2 &   0.1135 \\
2 & 1 & 2 & 1 &   0.1135 \\
2 & 2 & 1 & 1 &  -0.1196 \\
3 & 1 & 1 & 3 &   1.0000 \\
3 & 1 & 3 & 1 &   1.0000 \\
3 & 3 & 1 & 1 &   0.7016 \\
\hline
\end{tabular}
\caption{base 9}
\end{subtable}
\caption{ orthotropic elastic modulus: 9 bases }
\label{tab:orthotropic elastic modulus}
\end{table}
\afterpage{\clearpage}

\clearpage
\setcounter{equation}{0}
\setcounter{table}{0}
\section{Components of fourthorder structure tensors for common symmetry groups} \label{app:4th_order}
\normalsize

Using \tref{tab:crystal_groups}b we find the following bases for the crystal subgroups.
We mark the non-exclusive basis elements with {\it italics}.
We omit reporting the trigonal basis with 9 bases for brevity due to the number of components per basis element.

\begin{table}
\centering
\begin{subtable}[c]{0.3\textwidth}
\rm
\begin{tabular}{|cccc|r|}
\hline
1 & 1 & 1 & 1 &   1.0000 \\
\hline
\end{tabular}
\caption{base 1}
\end{subtable}
\begin{subtable}[c]{0.3\textwidth}
\rm
\begin{tabular}{|cccc|r|}
\hline
3 & 3 & 1 & 1 &   1.0000 \\
\hline
\end{tabular}
\caption{base 2}
\end{subtable}
\begin{subtable}[c]{0.3\textwidth}
\rm
\begin{tabular}{|cccc|r|}
\hline
3 & 3 & 2 & 2 &   1.0000 \\
\hline
\end{tabular}
\caption{base 3}
\end{subtable}
\begin{subtable}[c]{0.3\textwidth}
\it
\begin{tabular}{|cccc|r|}
\hline
3 & 3 & 3 & 3 &   1.0000 \\
\hline
\end{tabular}
\caption{base 4}
\end{subtable}
\begin{subtable}[c]{0.3\textwidth}
\rm
\begin{tabular}{|cccc|r|}
\hline
3 & 2 & 3 & 2 &   1.0000 \\
\hline
\end{tabular}
\caption{base 5}
\end{subtable}
\begin{subtable}[c]{0.3\textwidth}
\rm
\begin{tabular}{|cccc|r|}
\hline
3 & 1 & 1 & 3 &   1.0000 \\
\hline
\end{tabular}
\caption{base 6}
\end{subtable}
\begin{subtable}[c]{0.3\textwidth}
\rm
\begin{tabular}{|cccc|r|}
\hline
2 & 2 & 2 & 2 &   1.0000 \\
\hline
\end{tabular}
\caption{base 7}
\end{subtable}
\begin{subtable}[c]{0.3\textwidth}
\rm
\begin{tabular}{|cccc|r|}
\hline
3 & 1 & 3 & 1 &  -0.1485 \\
3 & 2 & 2 & 3 &   1.0000 \\
\hline
\end{tabular}
\caption{base 8}
\end{subtable}
\begin{subtable}[c]{0.3\textwidth}
\rm
\begin{tabular}{|cccc|r|}
\hline
2 & 3 & 2 & 3 &   1.0000 \\
2 & 3 & 3 & 2 &   0.0349 \\
\hline
\end{tabular}
\caption{base 9}
\end{subtable}
\begin{subtable}[c]{0.3\textwidth}
\rm
\begin{tabular}{|cccc|r|}
\hline
3 & 1 & 3 & 1 &   1.0000 \\
3 & 2 & 2 & 3 &   0.1485 \\
\hline
\end{tabular}
\caption{base 10}
\end{subtable}
\begin{subtable}[c]{0.3\textwidth}
\rm
\begin{tabular}{|cccc|r|}
\hline
2 & 3 & 2 & 3 &  -0.0349 \\
2 & 3 & 3 & 2 &   1.0000 \\
\hline
\end{tabular}
\caption{base 11}
\end{subtable}
\begin{subtable}[c]{0.3\textwidth}
\rm
\begin{tabular}{|cccc|r|}
\hline
1 & 3 & 1 & 3 &   1.0000 \\
1 & 3 & 3 & 1 &   0.2232 \\
2 & 2 & 3 & 3 &   0.0002 \\
\hline
\end{tabular}
\caption{base 12}
\end{subtable}
\begin{subtable}[c]{0.3\textwidth}
\rm
\begin{tabular}{|cccc|r|}
\hline
1 & 3 & 1 & 3 &  -0.2232 \\
1 & 3 & 3 & 1 &   1.0000 \\
2 & 1 & 1 & 2 &  -0.5136 \\
2 & 1 & 2 & 1 &   0.2750 \\
2 & 2 & 1 & 1 &   0.2386 \\
\hline
\end{tabular}
\caption{base 13}
\end{subtable}
\begin{subtable}[c]{0.3\textwidth}
\rm
\begin{tabular}{|cccc|r|}
\hline
1 & 3 & 1 & 3 &  -0.1669 \\
1 & 3 & 3 & 1 &   0.7480 \\
2 & 1 & 1 & 2 &   1.0000 \\
2 & 1 & 2 & 1 &  -0.9058 \\
2 & 2 & 1 & 1 &  -0.0942 \\
\hline
\end{tabular}
\caption{base 14}
\end{subtable}
\begin{subtable}[c]{0.3\textwidth}
\rm
\begin{tabular}{|cccc|r|}
\hline
1 & 3 & 1 & 3 &   0.0485 \\
1 & 3 & 3 & 1 &  -0.2179 \\
2 & 1 & 1 & 2 &  -0.3361 \\
2 & 1 & 2 & 1 &  -0.6639 \\
2 & 2 & 1 & 1 &   1.0000 \\
2 & 2 & 3 & 3 &   0.4263 \\
\hline
\end{tabular}
\caption{base 15}
\end{subtable}
\begin{subtable}[c]{0.3\textwidth}
\rm
\begin{tabular}{|cccc|r|}
\hline
1 & 3 & 1 & 3 &  -0.0131 \\
1 & 3 & 3 & 1 &   0.0579 \\
2 & 1 & 1 & 2 &   0.0893 \\
2 & 1 & 2 & 1 &   0.1765 \\
2 & 2 & 1 & 1 &  -0.2658 \\
2 & 2 & 3 & 3 &   1.0000 \\
\hline
\end{tabular}
\caption{base 16}
\end{subtable}
\begin{subtable}[c]{0.3\textwidth}
\it
\begin{tabular}{|cccc|r|}
\hline
1 & 1 & 2 & 2 &   1.0000 \\
1 & 2 & 1 & 2 &   1.0000 \\
1 & 2 & 2 & 1 &   1.0000 \\
2 & 1 & 1 & 2 &   1.0000 \\
2 & 1 & 2 & 1 &   1.0000 \\
2 & 2 & 1 & 1 &   1.0000 \\
\hline
\end{tabular}
\caption{base 17}
\end{subtable}
\begin{subtable}[c]{0.3\textwidth}
\rm
\begin{tabular}{|cccc|r|}
\hline
1 & 1 & 2 & 2 &   1.0000 \\
1 & 1 & 3 & 3 &  -0.5132 \\
1 & 2 & 1 & 2 &  -0.7430 \\
1 & 2 & 2 & 1 &   0.2327 \\
2 & 1 & 1 & 2 &  -0.1632 \\
2 & 1 & 2 & 1 &  -0.1632 \\
2 & 2 & 1 & 1 &  -0.1632 \\
\hline
\end{tabular}
\caption{base 18}
\end{subtable}
\begin{subtable}[c]{0.3\textwidth}
\rm
\begin{tabular}{|cccc|r|}
\hline
1 & 1 & 2 & 2 &   0.4238 \\
1 & 1 & 3 & 3 &  -0.6410 \\
1 & 2 & 1 & 2 &   1.0000 \\
1 & 2 & 2 & 1 &  -0.6116 \\
2 & 1 & 1 & 2 &  -0.2707 \\
2 & 1 & 2 & 1 &  -0.2707 \\
2 & 2 & 1 & 1 &  -0.2707 \\
\hline
\end{tabular}
\caption{base 19}
\end{subtable}
\begin{subtable}[c]{0.3\textwidth}
\rm
\begin{tabular}{|cccc|r|}
\hline
1 & 1 & 2 & 2 &   0.5381 \\
1 & 1 & 3 & 3 &   1.0000 \\
1 & 2 & 1 & 2 &   0.2206 \\
1 & 2 & 2 & 1 &   0.0382 \\
2 & 1 & 1 & 2 &  -0.2656 \\
2 & 1 & 2 & 1 &  -0.2656 \\
2 & 2 & 1 & 1 &  -0.2656 \\
\hline
\end{tabular}
\caption{base 20}
\end{subtable}
\begin{subtable}[c]{0.3\textwidth}
\rm
\begin{tabular}{|cccc|r|}
\hline
1 & 1 & 2 & 2 &  -0.2851 \\
1 & 1 & 3 & 3 &  -0.2255 \\
1 & 2 & 1 & 2 &   0.3103 \\
1 & 2 & 2 & 1 &   1.0000 \\
2 & 1 & 1 & 2 &  -0.3417 \\
2 & 1 & 2 & 1 &  -0.3417 \\
2 & 2 & 1 & 1 &  -0.3417 \\
\hline
\end{tabular}
\caption{base 21}
\end{subtable}
\caption{ orthotropic: 21 bases }
\label{tab:orthotropic}
\end{table}
\afterpage{\clearpage}

\begin{table}
\centering
\begin{subtable}[c]{0.3\textwidth}
\it
\begin{tabular}{|cccc|r|}
\hline
3 & 3 & 3 & 3 &   1.0000 \\
\hline
\end{tabular}
\caption{base 1}
\end{subtable}
\begin{subtable}[c]{0.3\textwidth}
\it
\begin{tabular}{|cccc|r|}
\hline
3 & 3 & 1 & 1 &   1.0000 \\
3 & 3 & 2 & 2 &   1.0000 \\
\hline
\end{tabular}
\caption{base 2}
\end{subtable}
\begin{subtable}[c]{0.3\textwidth}
\rm
\begin{tabular}{|cccc|r|}
\hline
1 & 1 & 1 & 1 &   1.0000 \\
2 & 2 & 2 & 2 &   1.0000 \\
\hline
\end{tabular}
\caption{base 3}
\end{subtable}
\begin{subtable}[c]{0.3\textwidth}
\it
\begin{tabular}{|cccc|r|}
\hline
3 & 1 & 1 & 3 &   1.0000 \\
3 & 1 & 3 & 1 &  -0.0110 \\
3 & 2 & 2 & 3 &   1.0000 \\
3 & 2 & 3 & 2 &  -0.0110 \\
\hline
\end{tabular}
\caption{base 4}
\end{subtable}
\begin{subtable}[c]{0.3\textwidth}
\it
\begin{tabular}{|cccc|r|}
\hline
3 & 1 & 1 & 3 &   0.0110 \\
3 & 1 & 3 & 1 &   1.0000 \\
3 & 2 & 2 & 3 &   0.0110 \\
3 & 2 & 3 & 2 &   1.0000 \\
\hline
\end{tabular}
\caption{base 5}
\end{subtable}
\begin{subtable}[c]{0.3\textwidth}
\rm
\begin{tabular}{|cccc|r|}
\hline
1 & 1 & 2 & 2 &   0.2101 \\
1 & 1 & 3 & 3 &  -0.2240 \\
1 & 2 & 1 & 2 &   1.0000 \\
1 & 2 & 2 & 1 &  -0.3943 \\
2 & 1 & 1 & 2 &  -0.3943 \\
2 & 1 & 2 & 1 &   1.0000 \\
2 & 2 & 1 & 1 &   0.2101 \\
2 & 2 & 3 & 3 &  -0.2240 \\
\hline
\end{tabular}
\caption{base 6}
\end{subtable}
\begin{subtable}[c]{0.3\textwidth}
\rm
\begin{tabular}{|cccc|r|}
\hline
1 & 1 & 2 & 2 &   0.3850 \\
1 & 1 & 3 & 3 &  -0.3101 \\
1 & 2 & 1 & 2 &  -0.2354 \\
1 & 2 & 2 & 1 &  -0.2156 \\
1 & 3 & 1 & 3 &   1.0000 \\
1 & 3 & 3 & 1 &   0.7316 \\
2 & 1 & 1 & 2 &  -0.2156 \\
2 & 1 & 2 & 1 &  -0.2354 \\
2 & 2 & 1 & 1 &   0.3850 \\
2 & 2 & 3 & 3 &  -0.3101 \\
2 & 3 & 2 & 3 &   1.0000 \\
2 & 3 & 3 & 2 &   0.7316 \\
\hline
\end{tabular}
\caption{base 7}
\end{subtable}
\begin{subtable}[c]{0.3\textwidth}
\rm
\begin{tabular}{|cccc|r|}
\hline
1 & 1 & 2 & 2 &  -0.3595 \\
1 & 1 & 3 & 3 &   1.0000 \\
1 & 2 & 1 & 2 &   0.3055 \\
1 & 2 & 2 & 1 &   0.0151 \\
1 & 3 & 1 & 3 &  -0.0324 \\
1 & 3 & 3 & 1 &   0.7602 \\
2 & 1 & 1 & 2 &   0.0151 \\
2 & 1 & 2 & 1 &   0.3055 \\
2 & 2 & 1 & 1 &  -0.3595 \\
2 & 2 & 3 & 3 &   1.0000 \\
2 & 3 & 2 & 3 &  -0.0324 \\
2 & 3 & 3 & 2 &   0.7602 \\
\hline
\end{tabular}
\caption{base 8}
\end{subtable}
\begin{subtable}[c]{0.3\textwidth}
\rm
\begin{tabular}{|cccc|r|}
\hline
1 & 1 & 2 & 2 &  -0.8005 \\
1 & 1 & 3 & 3 &  -0.8166 \\
1 & 2 & 1 & 2 &   0.3795 \\
1 & 2 & 2 & 1 &   1.0000 \\
1 & 3 & 1 & 3 &  -0.0223 \\
1 & 3 & 3 & 1 &   0.5224 \\
2 & 1 & 1 & 2 &   1.0000 \\
2 & 1 & 2 & 1 &   0.3795 \\
2 & 2 & 1 & 1 &  -0.8005 \\
2 & 2 & 3 & 3 &  -0.8166 \\
2 & 3 & 2 & 3 &  -0.0223 \\
2 & 3 & 3 & 2 &   0.5224 \\
\hline
\end{tabular}
\caption{base 9}
\end{subtable}
\begin{subtable}[c]{0.3\textwidth}
\rm
\begin{tabular}{|cccc|r|}
\hline
1 & 1 & 2 & 2 &   1.0000 \\
1 & 1 & 3 & 3 &   0.3182 \\
1 & 2 & 1 & 2 &   0.2551 \\
1 & 2 & 2 & 1 &   0.9990 \\
1 & 3 & 1 & 3 &   0.0377 \\
1 & 3 & 3 & 1 &  -0.0663 \\
2 & 1 & 1 & 2 &   0.9990 \\
2 & 1 & 2 & 1 &   0.2551 \\
2 & 2 & 1 & 1 &   1.0000 \\
2 & 2 & 3 & 3 &   0.3182 \\
2 & 3 & 2 & 3 &   0.0377 \\
2 & 3 & 3 & 2 &  -0.0663 \\
\hline
\end{tabular}
\caption{base 10}
\end{subtable}
\begin{subtable}[c]{0.3\textwidth}
\rm
\begin{tabular}{|cccc|r|}
\hline
1 & 1 & 2 & 2 &  -0.5031 \\
1 & 1 & 3 & 3 &   0.3653 \\
1 & 2 & 1 & 2 &   0.2768 \\
1 & 2 & 2 & 1 &   0.2263 \\
1 & 3 & 1 & 3 &   1.0000 \\
1 & 3 & 3 & 1 &  -0.7915 \\
2 & 1 & 1 & 2 &   0.2263 \\
2 & 1 & 2 & 1 &   0.2768 \\
2 & 2 & 1 & 1 &  -0.5031 \\
2 & 2 & 3 & 3 &   0.3653 \\
2 & 3 & 2 & 3 &   1.0000 \\
2 & 3 & 3 & 2 &  -0.7915 \\
\hline
\end{tabular}
\caption{base 11}
\end{subtable}
\caption{ tetragonal: 11 bases }
\label{tab:tetragonal}
\end{table}
\afterpage{\clearpage}

\begin{table}
\centering
\begin{subtable}[c]{0.3\textwidth}
\rm
\begin{tabular}{|cccc|r|}
\hline
1 & 1 & 1 & 1 &   1.0000 \\
2 & 2 & 2 & 2 &   1.0000 \\
3 & 3 & 3 & 3 &   1.0000 \\
\hline
\end{tabular}
\caption{base 1}
\end{subtable}
\begin{subtable}[c]{0.3\textwidth}
\rm
\begin{tabular}{|cccc|r|}
\hline
1 & 1 & 2 & 2 &  -0.2808 \\
1 & 1 & 3 & 3 &  -0.2808 \\
1 & 2 & 1 & 2 &   0.2779 \\
1 & 2 & 2 & 1 &   1.0000 \\
1 & 3 & 1 & 3 &   0.2779 \\
1 & 3 & 3 & 1 &   1.0000 \\
2 & 1 & 1 & 2 &   1.0000 \\
2 & 1 & 2 & 1 &   0.2779 \\
2 & 2 & 1 & 1 &  -0.2808 \\
2 & 2 & 3 & 3 &  -0.2808 \\
2 & 3 & 2 & 3 &   0.2779 \\
2 & 3 & 3 & 2 &   1.0000 \\
3 & 1 & 1 & 3 &   1.0000 \\
3 & 1 & 3 & 1 &   0.2779 \\
3 & 2 & 2 & 3 &   1.0000 \\
3 & 2 & 3 & 2 &   0.2779 \\
3 & 3 & 1 & 1 &  -0.2808 \\
3 & 3 & 2 & 2 &  -0.2808 \\
\hline
\end{tabular}
\caption{base 2}
\end{subtable}
\begin{subtable}[c]{0.3\textwidth}
\rm
\begin{tabular}{|cccc|r|}
\hline
1 & 1 & 2 & 2 &   1.0000 \\
1 & 1 & 3 & 3 &   1.0000 \\
1 & 2 & 1 & 2 &   0.2076 \\
1 & 2 & 2 & 1 &   0.2231 \\
1 & 3 & 1 & 3 &   0.2076 \\
1 & 3 & 3 & 1 &   0.2231 \\
2 & 1 & 1 & 2 &   0.2231 \\
2 & 1 & 2 & 1 &   0.2076 \\
2 & 2 & 1 & 1 &   1.0000 \\
2 & 2 & 3 & 3 &   1.0000 \\
2 & 3 & 2 & 3 &   0.2076 \\
2 & 3 & 3 & 2 &   0.2231 \\
3 & 1 & 1 & 3 &   0.2231 \\
3 & 1 & 3 & 1 &   0.2076 \\
3 & 2 & 2 & 3 &   0.2231 \\
3 & 2 & 3 & 2 &   0.2076 \\
3 & 3 & 1 & 1 &   1.0000 \\
3 & 3 & 2 & 2 &   1.0000 \\
\hline
\end{tabular}
\caption{base 3}
\end{subtable}
\begin{subtable}[c]{0.3\textwidth}
\rm
\begin{tabular}{|cccc|r|}
\hline
1 & 1 & 2 & 2 &  -0.1370 \\
1 & 1 & 3 & 3 &  -0.1370 \\
1 & 2 & 1 & 2 &   1.0000 \\
1 & 2 & 2 & 1 &  -0.3164 \\
1 & 3 & 1 & 3 &   1.0000 \\
1 & 3 & 3 & 1 &  -0.3164 \\
2 & 1 & 1 & 2 &  -0.3164 \\
2 & 1 & 2 & 1 &   1.0000 \\
2 & 2 & 1 & 1 &  -0.1370 \\
2 & 2 & 3 & 3 &  -0.1370 \\
2 & 3 & 2 & 3 &   1.0000 \\
2 & 3 & 3 & 2 &  -0.3164 \\
3 & 1 & 1 & 3 &  -0.3164 \\
3 & 1 & 3 & 1 &   1.0000 \\
3 & 2 & 2 & 3 &  -0.3164 \\
3 & 2 & 3 & 2 &   1.0000 \\
3 & 3 & 1 & 1 &  -0.1370 \\
3 & 3 & 2 & 2 &  -0.1370 \\
\hline
\end{tabular}
\caption{base 4}
\end{subtable}
\caption{ cubic: 4 bases }
\label{tab:cubic}
\end{table}
\afterpage{\clearpage}

\begin{table}
\centering
\begin{subtable}[c]{0.3\textwidth}
\rm
\begin{tabular}{|cccc|r|}
\hline
3 & 3 & 3 & 3 &   1.0000 \\
\hline
\end{tabular}
\caption{base 1}
\end{subtable}
\begin{subtable}[c]{0.3\textwidth}
\rm
\begin{tabular}{|cccc|r|}
\hline
3 & 3 & 1 & 1 &   1.0000 \\
3 & 3 & 2 & 2 &   1.0000 \\
\hline
\end{tabular}
\caption{base 2}
\end{subtable}
\begin{subtable}[c]{0.3\textwidth}
\rm
\begin{tabular}{|cccc|r|}
\hline
3 & 1 & 1 & 3 &   0.9785 \\
3 & 1 & 3 & 1 &   1.0000 \\
3 & 2 & 2 & 3 &   0.9785 \\
3 & 2 & 3 & 2 &   1.0000 \\
\hline
\end{tabular}
\caption{base 3}
\end{subtable}
\begin{subtable}[c]{0.3\textwidth}
\rm
\begin{tabular}{|cccc|r|}
\hline
3 & 1 & 1 & 3 &   1.0000 \\
3 & 1 & 3 & 1 &  -0.9785 \\
3 & 2 & 2 & 3 &   1.0000 \\
3 & 2 & 3 & 2 &  -0.9785 \\
\hline
\end{tabular}
\caption{base 4}
\end{subtable}
\begin{subtable}[c]{0.3\textwidth}
\rm
\begin{tabular}{|cccc|r|}
\hline
1 & 1 & 1 & 1 &  -0.0861 \\
1 & 1 & 2 & 2 &   1.0000 \\
1 & 1 & 3 & 3 &  -0.3877 \\
1 & 2 & 1 & 2 &  -0.4846 \\
1 & 2 & 2 & 1 &  -0.6015 \\
1 & 3 & 1 & 3 &  -0.1710 \\
1 & 3 & 3 & 1 &   0.4508 \\
2 & 1 & 1 & 2 &  -0.6015 \\
2 & 1 & 2 & 1 &  -0.4846 \\
2 & 2 & 1 & 1 &   1.0000 \\
2 & 2 & 2 & 2 &  -0.0861 \\
2 & 2 & 3 & 3 &  -0.3877 \\
2 & 3 & 2 & 3 &  -0.1710 \\
2 & 3 & 3 & 2 &   0.4508 \\
\hline
\end{tabular}
\caption{base 5}
\end{subtable}
\begin{subtable}[c]{0.3\textwidth}
\rm
\begin{tabular}{|cccc|r|}
\hline
1 & 1 & 1 & 1 &   0.8918 \\
1 & 1 & 2 & 2 &   0.0665 \\
1 & 1 & 3 & 3 &  -0.4763 \\
1 & 2 & 1 & 2 &   1.0000 \\
1 & 2 & 2 & 1 &  -0.1747 \\
1 & 3 & 1 & 3 &  -0.4138 \\
1 & 3 & 3 & 1 &   0.2982 \\
2 & 1 & 1 & 2 &  -0.1747 \\
2 & 1 & 2 & 1 &   1.0000 \\
2 & 2 & 1 & 1 &   0.0665 \\
2 & 2 & 2 & 2 &   0.8918 \\
2 & 2 & 3 & 3 &  -0.4763 \\
2 & 3 & 2 & 3 &  -0.4138 \\
2 & 3 & 3 & 2 &   0.2982 \\
\hline
\end{tabular}
\caption{base 6}
\end{subtable}
\begin{subtable}[c]{0.3\textwidth}
\rm
\begin{tabular}{|cccc|r|}
\hline
1 & 1 & 1 & 1 &  -0.0195 \\
1 & 1 & 2 & 2 &  -0.1637 \\
1 & 1 & 3 & 3 &   0.3878 \\
1 & 2 & 1 & 2 &   0.0733 \\
1 & 2 & 2 & 1 &   0.0709 \\
1 & 3 & 1 & 3 &   0.3530 \\
1 & 3 & 3 & 1 &   1.0000 \\
2 & 1 & 1 & 2 &   0.0709 \\
2 & 1 & 2 & 1 &   0.0733 \\
2 & 2 & 1 & 1 &  -0.1637 \\
2 & 2 & 2 & 2 &  -0.0195 \\
2 & 2 & 3 & 3 &   0.3878 \\
2 & 3 & 2 & 3 &   0.3530 \\
2 & 3 & 3 & 2 &   1.0000 \\
\hline
\end{tabular}
\caption{base 7}
\end{subtable}
\begin{subtable}[c]{0.3\textwidth}
\rm
\begin{tabular}{|cccc|r|}
\hline
1 & 1 & 1 & 1 &   0.5924 \\
1 & 1 & 2 & 2 &   0.1887 \\
1 & 1 & 3 & 3 &  -0.1309 \\
1 & 2 & 1 & 2 &  -0.5964 \\
1 & 2 & 2 & 1 &   1.0000 \\
1 & 3 & 1 & 3 &  -0.2856 \\
1 & 3 & 3 & 1 &   0.1668 \\
2 & 1 & 1 & 2 &   1.0000 \\
2 & 1 & 2 & 1 &  -0.5964 \\
2 & 2 & 1 & 1 &   0.1887 \\
2 & 2 & 2 & 2 &   0.5924 \\
2 & 2 & 3 & 3 &  -0.1309 \\
2 & 3 & 2 & 3 &  -0.2856 \\
2 & 3 & 3 & 2 &   0.1668 \\
\hline
\end{tabular}
\caption{base 8}
\end{subtable}
\begin{subtable}[c]{0.3\textwidth}
\rm
\begin{tabular}{|cccc|r|}
\hline
1 & 1 & 1 & 1 &   0.9218 \\
1 & 1 & 2 & 2 &   0.8923 \\
1 & 1 & 3 & 3 &   1.0000 \\
1 & 2 & 1 & 2 &   0.1493 \\
1 & 2 & 2 & 1 &  -0.1198 \\
1 & 3 & 1 & 3 &   0.9782 \\
1 & 3 & 3 & 1 &  -0.5715 \\
2 & 1 & 1 & 2 &  -0.1198 \\
2 & 1 & 2 & 1 &   0.1493 \\
2 & 2 & 1 & 1 &   0.8923 \\
2 & 2 & 2 & 2 &   0.9218 \\
2 & 2 & 3 & 3 &   1.0000 \\
2 & 3 & 2 & 3 &   0.9782 \\
2 & 3 & 3 & 2 &  -0.5715 \\
\hline
\end{tabular}
\caption{base 9}
\end{subtable}
\begin{subtable}[c]{0.3\textwidth}
\rm
\begin{tabular}{|cccc|r|}
\hline
1 & 1 & 1 & 1 &   0.0381 \\
1 & 1 & 2 & 2 &  -0.0927 \\
1 & 1 & 3 & 3 &  -0.9207 \\
1 & 2 & 1 & 2 &  -0.0203 \\
1 & 2 & 2 & 1 &   0.1511 \\
1 & 3 & 1 & 3 &   1.0000 \\
1 & 3 & 3 & 1 &  -0.0196 \\
2 & 1 & 1 & 2 &   0.1511 \\
2 & 1 & 2 & 1 &  -0.0203 \\
2 & 2 & 1 & 1 &  -0.0927 \\
2 & 2 & 2 & 2 &   0.0381 \\
2 & 2 & 3 & 3 &  -0.9207 \\
2 & 3 & 2 & 3 &   1.0000 \\
2 & 3 & 3 & 2 &  -0.0196 \\
\hline
\end{tabular}
\caption{base 10}
\end{subtable}
\caption{ transverse: 10 bases }
\label{tab:transverse}
\end{table}
\afterpage{\clearpage}

\begin{table}
\centering
\begin{subtable}[c]{0.3\textwidth}
\rm
\begin{tabular}{|cccc|r|}
\hline
1 & 1 & 1 & 1 &   1.0000 \\
1 & 1 & 2 & 2 &   0.0613 \\
1 & 1 & 3 & 3 &   0.0613 \\
1 & 2 & 1 & 2 &   0.5825 \\
1 & 2 & 2 & 1 &   0.3563 \\
1 & 3 & 1 & 3 &   0.5825 \\
1 & 3 & 3 & 1 &   0.3563 \\
2 & 1 & 1 & 2 &   0.3563 \\
2 & 1 & 2 & 1 &   0.5825 \\
2 & 2 & 1 & 1 &   0.0613 \\
2 & 2 & 2 & 2 &   1.0000 \\
2 & 2 & 3 & 3 &   0.0613 \\
2 & 3 & 2 & 3 &   0.5825 \\
2 & 3 & 3 & 2 &   0.3563 \\
3 & 1 & 1 & 3 &   0.3563 \\
3 & 1 & 3 & 1 &   0.5825 \\
3 & 2 & 2 & 3 &   0.3563 \\
3 & 2 & 3 & 2 &   0.5825 \\
3 & 3 & 1 & 1 &   0.0613 \\
3 & 3 & 2 & 2 &   0.0613 \\
3 & 3 & 3 & 3 &   1.0000 \\
\hline
\end{tabular}
\caption{base 1}
\end{subtable}
\begin{subtable}[c]{0.3\textwidth}
\rm
\begin{tabular}{|cccc|r|}
\hline
1 & 1 & 1 & 1 &   0.2349 \\
1 & 1 & 2 & 2 &   0.0539 \\
1 & 1 & 3 & 3 &   0.0539 \\
1 & 2 & 1 & 2 &  -0.8190 \\
1 & 2 & 2 & 1 &   1.0000 \\
1 & 3 & 1 & 3 &  -0.8190 \\
1 & 3 & 3 & 1 &   1.0000 \\
2 & 1 & 1 & 2 &   1.0000 \\
2 & 1 & 2 & 1 &  -0.8190 \\
2 & 2 & 1 & 1 &   0.0539 \\
2 & 2 & 2 & 2 &   0.2349 \\
2 & 2 & 3 & 3 &   0.0539 \\
2 & 3 & 2 & 3 &  -0.8190 \\
2 & 3 & 3 & 2 &   1.0000 \\
3 & 1 & 1 & 3 &   1.0000 \\
3 & 1 & 3 & 1 &  -0.8190 \\
3 & 2 & 2 & 3 &   1.0000 \\
3 & 2 & 3 & 2 &  -0.8190 \\
3 & 3 & 1 & 1 &   0.0539 \\
3 & 3 & 2 & 2 &   0.0539 \\
3 & 3 & 3 & 3 &   0.2349 \\
\hline
\end{tabular}
\caption{base 2}
\end{subtable}
\begin{subtable}[c]{0.3\textwidth}
\rm
\begin{tabular}{|cccc|r|}
\hline
1 & 1 & 1 & 1 &   0.4146 \\
1 & 1 & 2 & 2 &   1.0000 \\
1 & 1 & 3 & 3 &   1.0000 \\
1 & 2 & 1 & 2 &  -0.2654 \\
1 & 2 & 2 & 1 &  -0.3200 \\
1 & 3 & 1 & 3 &  -0.2654 \\
1 & 3 & 3 & 1 &  -0.3200 \\
2 & 1 & 1 & 2 &  -0.3200 \\
2 & 1 & 2 & 1 &  -0.2654 \\
2 & 2 & 1 & 1 &   1.0000 \\
2 & 2 & 2 & 2 &   0.4146 \\
2 & 2 & 3 & 3 &   1.0000 \\
2 & 3 & 2 & 3 &  -0.2654 \\
2 & 3 & 3 & 2 &  -0.3200 \\
3 & 1 & 1 & 3 &  -0.3200 \\
3 & 1 & 3 & 1 &  -0.2654 \\
3 & 2 & 2 & 3 &  -0.3200 \\
3 & 2 & 3 & 2 &  -0.2654 \\
3 & 3 & 1 & 1 &   1.0000 \\
3 & 3 & 2 & 2 &   1.0000 \\
3 & 3 & 3 & 3 &   0.4146 \\
\hline
\end{tabular}
\caption{base 3}
\end{subtable}
\caption{ isotropic: 3 bases }
\label{tab:isotropic}
\end{table}
\afterpage{\clearpage}
\end{document}